\documentclass[iop,apj]{emulateapj}
\usepackage{amsmath,amssymb,amstext}
\usepackage{graphicx}
\usepackage{color,soul}

\usepackage[breaklinks,colorlinks,citecolor=blue,linkcolor=magenta]{hyperref}

\shorttitle{RM Synthesis for Probing Intervening Matter}
\shortauthors{Kim et al.}

\begin{document}

\title{Faraday Rotation Measure Synthesis of intermediate redshift quasars as a probe of intervening matter}
\author{Kwang Seong Kim\altaffilmark{1}}
\author{Simon J. Lilly\altaffilmark{1}}
\author{Francesco Miniati\altaffilmark{1}}
\author{Martin L. Bernet\altaffilmark{1}}
\author{Rainer Beck\altaffilmark{2}}
\author{Shane P. O'Sullivan\altaffilmark{3}\altaffilmark{4}}
\author{Bryan M. Gaensler\altaffilmark{5}\altaffilmark{4}}
\email{kwangseong.kim@phys.ethz.ch}
\altaffiltext{1}{Physics Department, ETH Zurich, Wolfgang-Pauli-Strasse 27, 8093 Zurich, Switzerland}
\altaffiltext{2}{Max-Planck-Institut f\"ur Radioastronomie, Auf dem H\"ugel 69, 53121 Bonn, Germany}
\altaffiltext{3}{Instituto de Astronom\'ia, Universidad Nacional Aut\'onoma de M\'exico (UNAM),
  A.P. 70-264, 04510 M\'exico, D.F., Mexico}
\altaffiltext{4}{Sydney Institute for Astronomy, School of Physics, The University of Sydney, NSW
  2006, Australia}
\altaffiltext{5}{Dunlap Institute for Astronomy and Astrophysics, University of Toronto, Toronto, ON
  M5S 3H4, Canada}

\begin{abstract}
  There is evidence that magnetized material along the line of sight to distant quasars is
  detectable in the polarization properties of the background sources. The polarization properties
  appear to be correlated with the presence of intervening MgII absorption, which is thought to
  arise in outflowing material from star forming galaxies. In order to investigate this further, we
  have obtained high spectral resolution polarization measurements, with the VLA and ATCA, of a set
  of 49 unresolved quasars for which we have high quality optical spectra. These enable us to
  produce a Faraday Depth spectrum for each source, using Rotation Measure Synthesis. Our new
  independent radio data confirms that interveners are strongly associated with depolarization. We
  characterize the complexity of the Faraday Depth spectrum using a number of parameters and show
  how these are related, or not, to the depolarization and to the presence of MgII absorption along
  the line of sight. We argue that complexity and structure in the Faraday Depth distribution likely
  arise from both intervening material and intrinsically to the background source and attempt to
  separate these. We find that the strong radio depolarization effects associated with intervening
  material at redshifts out to $z \approx 1$ arise from inhomogeneous Faraday screens producing a
  dispersion in Rotation Measure across individual sources of around 10~rad/m$^2$. This is likely
  produced by disordered fields with strengths of at least $3\;\mu$G.
\end{abstract}

\maketitle

\section{Introduction}\label{sec:intro}
Cosmic magnetic fields are difficult to observe and difficult to treat theoretically, either
analytically or numerically.  Consequently fields on the scale of galaxies and larger are still
poorly understood and therefore often ignored in the context of the formation and evolution of
galaxies. However, detections of magnetic fields in a broad range of astrophysical objects have
provided indications for the ubiquitous nature of magnetic fields in the Universe.

Over the last few years, we have developed a line of research aimed at detecting and characterizing
magnetized material in and around normal galaxies at high redshift. The approach has been
to study compact radio-loud quasars and to search for correlations between
the presence of intervening strong MgII absorption in the optical spectra and the polarization
properties of the quasars, using the Faraday Rotation. Faraday Rotation describes the rotation of
the plane of polarization when polarized electromagnetic radiation traverses a region containing a
magnetized plasma.

The amount of rotation depends linearly on the square of the wavelength $\lambda^2$. The Rotation
Measure (RM) is defined as the gradient of the polarization angle $\chi$ against $\lambda^2$
\begin{equation}
\mathrm{RM} = \frac{\Delta\chi}{\Delta\lambda^2} \;.
\end{equation}
Assuming a simple case in which all radiation undergoes the same amount of rotation RM corresponds
to the Faraday Depth $\phi$ computed as (\cite{bur66})
\begin{equation}
  \phi(z_s) = 8.1 \times10^5\int_{z_s}^0\frac{n_e(z)B_{\parallel}(z)}{(1+z)^2}\frac{dl}{dz}dz \label{eq:RM}
\end{equation}
where $\phi$ is in units of $\mathrm{rad/m^{2}}$, the free electron number density $n_e$ in units of
$\mathrm{cm^{-3}}$, the magnetic field component along the line of sight $B_\parallel$ in units of
$\mathrm{G}$ and the comoving path increment per unit redshift $dl/dz$ in units of
$\mathrm{pc}$. Thus RM can be used to estimate the parallel component of the magnetic field along
the observed line of sight.

Unfortunately, the measurement of the RM gives no information about where along the line of sight
the Faraday Rotation is occurring, i.e. intrinsically to the radio source or its immediate
surroundings, during passage through an intervening system along the line of sight, or locally
within our own Galaxy.  If different parts of a source suffer different amounts of Faraday Rotation,
e.g. if there is a spread in $\phi$ caused by passing through an inhomogeneous foreground screen,
then because the polarization is a polar quantity, the net effect may be to reduce the overall
polarization of the source (\cite{bur66}, see also \cite{gar66}, \cite{sok98}).  These
depolarization effects can cause non-linearities in the slope of $\chi$ against $\lambda^2$ and make
the RM poorly defined.

\cite{kro08} presented evidence that the distribution of the apparent RM of a large sample of
radio quasars became broader with redshift. One possibility was that this was due to the increased
probability, as the redshift increases, that a given line of sight intersects intervening material.
To test this hypothesis, \cite{bern08} obtained high resolution optical spectra of a set of 76
quasars for which RM were available.  That analysis showed that, indeed, systems with strong MgII
absorption in their spectra had a broader distribution of RM than those without, and argued that
this was not being caused by a secondary correlation of the presence of MgII with
e.g. redshift. \cite{bern08} estimated that the lines of sight with strong MgII absorption were
suffering, statistically, an increased $|$RM$|$~$\approx40\;\mathrm{rad/m^2}$.  Applying a crude estimate
for the free electron column density led to an estimate of magnetic field strengths of order
10~$\mu G$ in the absorption systems, which are typically at redshift $z\approx 1$.

Furthermore by taking optical images of those quasars with MgII absorption, the absorbing systems
could be associated with individual galaxies (\cite{bern13}).  High values of RM were found only for
MgII systems that lay within 50 projected kpc.  These findings were very interesting in the context
of the results of \cite{bor11} who had mapped MgII absorption around galaxies at similar redshifts
and shown that most MgII absorption at these redshifts occurs in biconical bipolar regions extending
out to 50~kpc around star forming galaxies. Taken together they suggested that the observed magnetic
fields were being transported out of the galaxies by winds.  This would have implications for not
only the development of magnetic fields in galaxies but also for the origin of magnetic fields in
the circumgalactic and even intergalactic media (\cite{bha13}, \cite{shu06}).

Meanwhile new large catalogs of RM measurements became available, especially from 1.4 GHz surveys
(e.g. \cite{tay09}, see also \cite{far14}). \cite{bern12} compared the new RM measurements from
\cite{tay09}, which overlapped partly with the sample they have used (\cite{kro08}), and found that
there were significant differences. Furthermore, with the new RM data the excess RM associated with
lines of sight with MgII absorption could not be recovered.

The new RM samples were however measured at a relatively low frequency of 1.4 GHz, while the
previous RM measurements were at higher frequencies of 5~GHz. \cite{bern12} showed, using a simple
toy model that the presence of inhomogeneous screens could lead to substantial depolarization
towards lower frequencies which could introduce complex behavior into the wavelength-dependence of
the polarization angle and thus mask the intervener effect when RM measurements from low frequency
surveys were used.  This was later tested in \cite{bern13} who claimed that depolarization was
strongly associated with impact parameter for the MgII systems.

Later \cite{jos13} also used the RM catalog of \cite{ham12}, which is a crossmatch of the
\cite{tay09} catalog with several QSO redshift catalogs and found a marginal excess of RM associated
with MgII absorption. Also, \cite{far15} confirmed, with a much larger sample of around 600 objects
which they had compiled from their catalog (\cite{far14}), the association between excess RM and the
presence of MgII absorption along the line of sight. They also found that the association was only
present if only flat spectrum sources were considered. This could be because they will be smaller in
size and therefore have their radio and optical sight lines more closely aligned. \cite{bern08} did
not have this problem since their sample had been selected to be both compact and to have only small
offsets between radio and optical emission.

With the advent of new technology it has become possible to further improve the quality of the radio
data by measuring polarization at high spectral resolution over a long baseline in wavelength. In
particular, such data allows Faraday Rotation Measure Synthesis (RM Synthesis).  RM Synthesis
transforms the complex representation of the polarization structure
\begin{equation}
  \mathbf{P}(\lambda^2)=P(\lambda^2)e^{2i\chi(\lambda^2)} = Q(\lambda^2)+iU(\lambda^2)
\end{equation}
into the complex Faraday Depth distribution
\begin{equation}\label{eq:fdfourier}
  \mathbf{F}(\phi) = \frac{1}{\pi} \int_{-\infty}^\infty \mathbf{P}(\lambda^2) e^{-2i\phi\lambda^2}d\lambda^2
\end{equation}
where $P$ is the polarized flux density, $\chi$ the polarization angle and $Q$ and $U$ the
corresponding stokes parameters. $\mathbf{F}(\phi)$ is again a complex distribution and can be
written as
\begin{equation}
  \mathbf{F}(\phi)=F(\phi)e^{2i\psi(\phi)} = \tilde{Q}(\phi)+i\tilde{U}(\phi) \label{eq:fd}
\end{equation}
where the amplitude $F(\phi)$ will be referred to as the Faraday Depth (FD) distribution and
$\psi(\phi)$ as the ``initial phase'' distribution. $F(\phi)$ is also called Faraday Dispersion
Function or Faraday Spectrum in different literatures. $\tilde{Q}$ and $\tilde{U}$ are completely
analogous to $Q$ and $U$ but in $\phi$ space.

The FD distribution $F(\phi)$ quantifies how much linearly polarized flux density has been subject
to a certain Faraday Depth $\phi$.  The initial phase distribution $\psi(\phi)$ represents the
effective angle of polarization (before any Faraday rotation takes place) of the linearly polarized
emission that lies at a given Faraday Depth $\phi$.  By effective angle, we mean the angle of the
polar sum of the linearly polarized components.

A given source may exhibit a range of $\phi$ because one or, most likely, both of the following
conditions hold.  There must be either (a) different $\phi$ across the source (i.e in the plane of
the sky) within the telescope angular resolution due to variations across the emitting source or
across an intervening screen or (b) a contribution to $\phi$ along the line of sight \textit{within}
the emitting source.

\begin{deluxetable*}{lrrccccc}
  \tablewidth{0.9\textwidth}
  \tabletypesize{\small} \tablecaption{Observed sources\label{tbl:qso}}
  \tablehead{\multicolumn{1}{l}{Name} & \colhead{RA (J2000)} & \colhead{Dec. (J2000)} & \colhead{$z_{\mathrm{QSO}}$} &
    \colhead{MgII Abs.} & \colhead{$z_{\mathrm{Abs}}$} & \colhead{Instrument} & \colhead{Obs. block$^1$}
  }
  \startdata
  PKS0130-17                & 01:32:43.4   & -16:54:48      & 1.02                         & yes
                 & 0.51 & ATCA                 & A-SB1   \\
  PKS0135-247               & 01:37:38.3   & -24:30:54      & 0.84                         & yes
                 & 0.47 & ATCA                 & A-SB1   \\
  PKS0139-09                & 01:41:25.8   & -09:28:44      & 0.73                         & yes
                 & 0.50 & ATCA                 & A-SB1   \\
  3C057                     & 02:01:57.2   & -11:32:33      & 0.67                         & no
                 &      & ATCA                 & A-SB1   \\
  PKS0202-17                & 02:04:57.7   & -17:01:19      & 1.74                         & yes
                 & 0.52 & ATCA                 & A-SB1   \\
  PKS0332-403               & 03:34:13.7   & -40:08:25      & 1.45                         & yes
                 & 1.21 & ATCA                 & A-SB3   \\
  PKS0402-362               & 04:03:53.7   & -36:05:02      & 1.42                         & yes
                 & 0.80 & ATCA                 & A-SB2   \\
  PKS0422-380               & 04:24:42.4   & -37:56:22      & 0.78                         & no
                 &      & ATCA                 & A-SB3   \\
  PKS0426-380               & 04:28:40.4   & -37:56:20      & 1.11                         & yes
                 & 0.56 & ATCA                 & A-SB2   \\
  PKS0506-61                & 05:06:43.9   & -61:09:41      & 1.09                         & yes
                 & 0.92 & ATCA                 & A-SB3   \\
  PKS0839+18                & 08:42:05.1   & +18:35:42      & 1.27                         & yes
                 & 0.71 & VLA                  & V-SB1 \\
  4C+01.24                  & 09:09:10.1   & +01:21:36      & 1.02                         & yes
                 & 0.54 & VLA                  & V-SB1 \\
  4C+02.27                  & 09:35:18.2   & +02:04:16      & 0.65                         & no
                 &      & VLA                  & V-SB1 \\
  OK+186                    & 09:54:56.8   & +17:43:32      & 1.48                         & no
                 &      & VLA                  & V-SB1 \\
  4C+19.34                  & 10:24:44.8   & +19:12:21      & 0.83                         & yes
                 & 0.53 & VLA                  & V-SB1 \\
  4C+06.41                  & 10:41:17.2   & +06:10:17      & 1.27                         & yes
                 & 0.44 & VLA                  & V-SB1 \\
  3C245                     & 10:42:44.5   & +12:03:32      & 1.03                         & yes
                 & 0.66 & VLA                  & V-SB1 \\
  4C+20.24                  & 10:58:17.9   & +19:51:51      & 1.11                         & yes
                 & 0.86 & VLA                  & V-SB1 \\
  PKS1111+149               & 11:13:58.7   & +14:42:27      & 0.87                         & yes
                 & 0.65 & VLA                  & V-SB1 \\
  PKS1127-14                & 11:30:07.1   & -14:49:27      & 1.19                         & no
                 &      & ATCA                 & A-SB2 \\
  PKS1143-245               & 11:46:08.1   & -24:47:32      & 1.94                         & yes
                 & 1.25 & ATCA                 & A-SB3 \\
  PKS1157+014               & 11:59:44.8   & +01:12:07      & 1.99                         & yes
                 & 1.94 & VLA                  & V-SB1 \\
  4C+13.46                  & 12:13:32.1   & +13:07:20      & 1.14                         & yes
                 & 0.77 & VLA                  & V-SB1 \\
  4C-02.55                  & 12:32:00.0   & -02:24:05      & 1.05                         & yes
                 & 0.40 & VLA                  & V-SB1 \\
  PKS1244-255               & 12:46:46.9   & -25:47:48      & 0.63                         & yes
                 & 0.49 & ATCA                 & A-SB2 \\
  ON+187                    & 12:54:38.3   & +11:41:06      & 0.87                         & no
                 &      & VLA                  & V-SB3 \\
  4C-00.50                  & 13:19:38.7   & -00:49:41      & 0.89                         & no
                 &      & VLA                  & V-SB3 \\
  4C+19.44                  & 13:57:04.4   & +19:19:08      & 0.72                         & yes
                 & 0.46 & VLA                  & V-SB3 \\
  3C298                     & 14:19:08.2   & +06:28:35      & 1.44                         & no
                 &  & VLA                      & V-SB3 \\
  PKSB1419-272              & 14:22:49.0   & -27:27:56      & 0.99                         & yes
                 & 0.56 & VLA                  & V-SB3 \\
  OQ+135                    & 14:23:30.1   & +11:59:51      & 1.61                         & yes
                 & 1.36 & VLA                  & V-SB3 \\
  4C-05.62                  & 14:56:41.5   & -06:17:42      & 1.25                         & no
                 &      & VLA                  & V-SB3 \\
  4C-05.64                  & 15:10:53.6   & -05:43:07      & 1.19                         & yes
                 & 0.38 & VLA                  & V-SB3 \\
  4C+05.64                  & 15:50:35.3   & +05:27:11      & 1.42                         & no
                 &      & VLA                  & V-SB3 \\
  PKS1615+029               & 16:17:49.9   & +02:46:44      & 1.34                         & yes
                 & 0.53 & VLA                  & V-SB3 \\
  OW-174                    & 20:47:19.7   & -16:39:06      & 1.93                         & yes
                 & 1.33 & VLA                  & V-SB2 \\
  OX-325                    & 21:18:10.7   & -30:19:15      & 0.98                         & no
                 &  & VLA                      & V-SB2 \\
  PKS2134+004               & 21:36:38.6   & +00:41:55      & 1.94                         & yes
                 & 0.63 & VLA                  & V-SB2 \\
  OX-173                    & 21:46:23.0   & -15:25:44      & 0.70                         & no
                 &      & ATCA                 & A-SB1 \\
  4C+6.69                   & 21:48:05.4   & +06:57:39      & 0.99                         & yes
                 & 0.79 & VLA                  & V-SB2 \\
  OX-192                    & 21:58:06.3   & -15:01:09      & 0.67                         & yes
                 & 0.39 & ATCA                 & A-SB1 \\
  PKS2204-54                & 22:07:43.7   & -53:46:34      & 1.21                         & yes
                 & 0.69 & ATCA                 & A-SB1 \\
  4C-3.79                   & 22:18:52.0   & -03:35:37      & 0.90                         & no
                 &      & VLA                  & V-SB2 \\
  PKS2223-05                & 22:25:47.3   & -04:57:02      & 1.40                         & yes
                 & 0.85 & VLA                  & V-SB2 \\
  PKS2227-08                & 22:29:40.0   & -08:32:54      & 1.56                         & no
                 &  & ATCA                     & A-SB1 \\
  4C+11.69                  & 22:32:36.4   & +11:43:50      & 1.04                         & yes
                 & 0.74 & VLA                  & V-SB2 \\
  PKS2243-123               & 22:46:18.2   & -12:06:52      & 0.63                         & no
                 &      & ATCA                 & A-SB1 \\
  3C454.3                   & 22:53:57.7   & +16:08:53      & 0.86                         & no
                 &      & VLA                  & V-SB2 \\
  PKS2326-477               & 23:29:17.7   & -47:30:19      & 1.30                         & yes
                 & 0.43 & ATCA & A-SB1
\enddata
\tablenotetext{1}{See text for specifications.}
\end{deluxetable*}

RM Synthesis is a useful tool for studying Faraday Rotation effects, and depolarization
effects, for radio sources.  However, RM Synthesis requires dense measurements over a continuous
frequency band to get a reasonable coverage of $\mathbf{P}(\lambda^2)$ (see \cite{bec12} for a
summary of the Faraday Depth resolution of current and future radio telescopes). This has not been
available in earlier works dealing with intermediate redshift magnetic fields.  The goal of the
current paper has been to obtain the FD distribution $F(\phi)$ of a substantial fraction of the
radio sources for which we have high quality information on MgII absorption and which we used in
earlier papers (\cite{bern08}, \cite{bern10}, \cite{bern13}).

This paper is structured as follows.  In Section~\ref{sec:data}, we first describe our sample as
well as the new observations, data reduction and measurements. We also describe how the RM Synthesis
has been carried out and provide a brief review of the general interpretation of this relatively
unfamiliar type of data. In Section~\ref{sec:parameters} we introduce some parameters which quantify
structures in the polarization and in the FD distribution and which we will use in the later parts
of the paper. In Section~\ref{sec:gal} we fist look at whether these are correlated with Galactic
latitude which would indicate a potential Galactic origin. We then turn in Section~\ref{sec:results}
to examine the association with intervening MgII absorption and begin the development of the main
results of the paper. We first look at the analogues of the previous analyses in \cite{bern08}
regarding the distribution of overall RM and depolarization with the presence of MgII absorption.
We find a strong association of intervening MgII with various depolarization signatures. Turning to
examine the structure within the FD distribution $F(\phi)$, which we would expect to cause the
depolarization, we find surprisingly little correlation between this structure and MgII. This leads
us to explore the links between $F(\phi)$ and different manifestations of depolarization, and
enables us to identify those features in $F(\phi)$ which arise in intervening systems, and those
which are likely to be associated with the intrinsic properties of the sources. Finally in
Section~\ref{sec:discussion} we discuss the implications which arise from our results and in
Section~\ref{sec:summary} we summarize our findings. \newline\newline

\section{Data}\label{sec:data}

\subsection{Sample}

The sample of quasars is selected from \cite{bern08} and therefore has exactly the same MgII
information available for every quasar from VLT UVES spectroscopy. Some sources from \cite{bern08}
were not observed simply because of scheduling constraints at the telescopes. With few exceptions,
quasars with Dec $> 10^\circ$ were observed with the Karl G. Jansky Very Large Array (VLA) while
those south of this were observed with the Australia Telescope Compact Array (ATCA). 34 objects were
observed with VLA and 25 with ATCA. However, 4 objects observed with VLA and 6 objects observed with
ATCA are not considered further for this analysis. The 4 VLA objects which are thrown out are 3C208,
3C281, 4C-06.35 and PKS1424-11. For those objects it turned out that the core is too faint (and thus
too low in S/N) for our analysis. We believe that in prior observations the radio lobes have been
confused with the core. 6 ATCA objects were discarded because they are sufficiently resolved such
that a phase calibration has not been possible. The decision to discard those objects was taken
without any knowledge about their MgII absorption properties to avoid any bias in the results.

A list of the final set of 49 sources is provided in Table~\ref{tbl:qso} including their position,
their redshift, the redshift of their absorbers if present, and a log of observations. We declare an
MgII absorbing system to be present if an absorption line with rest frame equivalent width
$W_0\ge0.1$~\AA~is detected along the line of sight. This deviates slightly from the earlier
analysis in \cite{bern08} in which a cut at $W_0=0.3$~\AA~has been applied. We will discuss our
choice later in this paper. Consequently the sample with interveners contains 33 objects while the
sample with clean lines of sight contains 16~objects.  Similar results are found if attention is
restricted only to stronger sources, albeit at reduced significance.

\subsection{VLA Observations, data reduction and flux density measurements}

The objects were observed in three scheduling blocks (V-SB, cf. Table \ref{tbl:qso}) in
configuration A with a maximum baseline of 36.4 km.  V-SB1 was taken on 2014~March~4, V-SB2 on
2014~April~24 and V-SB3 on 2014~April~23 (Proposal ID: VLA/14A-144, PI: F. Miniati). The
observations were taken using the 8 bit sampler in L, S and C band, respectively, and cover with
full polarization the frequency range between approximately 1~and~6~GHz, implying a maximum
frequency resolution of 2~MHz in L~band and 1~MHz in S and C~band.

\begin{deluxetable}{ll}
  \tabletypesize{\small} \tablecaption{Phase Calibrators for the VLA sources}
  \tablewidth{0.4\textwidth}
  \tablecolumns{2}
  \tablehead{\multicolumn{1}{l}{Name} &\multicolumn{1}{l}{Phase Calibrators}} 
  \startdata
  PKS0839+18                & self \\
  4C+01.24                  & self \\
  4C+02.27                  & 4C+01.24 \\
  OK+186                    & self \\
  4C+19.34                  & OK+186 \\
  4C+06.41                  & self \\
  3C245                     & J1120+1420 \\
  4C+20.24                  & J1120+1420 \\
  PKS1111+149               & J1120+1420 \\
  PKS1157+014               & J1224+0330 \\
  4C+13.46                  & J1224+0330 \\
  4C-02.55                  & J1224+0330 \\
  ON+187                    & self \\
  4C-00.50                  & J1354-0206 \\
  4C+19.44                  & self \\
  3C298                     & J1415+1320 \\
  PKSB1419-272              & J1438-2204 \\
  OQ+135                    & 4C+19.44 \\
  4C-05.62                  & J1513-1012 \\
  4C-05.64                  & J1513-1012 \\
  4C+05.64                  & PKS1615+029 \\
  PKS1615+029               & self \\
  OW-174                    & J2110-1020/self$^1$ \\
  OX-325                    & J2138-2439 \\
  PKS2134+004               & self \\
  4C+6.69                   & self \\
  4C-3.79                   & J2212+0152/self$^1$ \\
  PKS2223-05                & J2212+0152/4C-3.79$^1$ \\
  4C+11.69                  & J2250+1419/self$^1$ \\
  3C454.3                   & J2250+1419/self$^1$
\enddata
\tablenotetext{1}{First calibrator used for L and S band and second calibrator used for C band.\newline}
\label{tbl:phasecal}
\end{deluxetable}

\begin{figure*}
  \centering
  \includegraphics[width=0.9\textwidth]{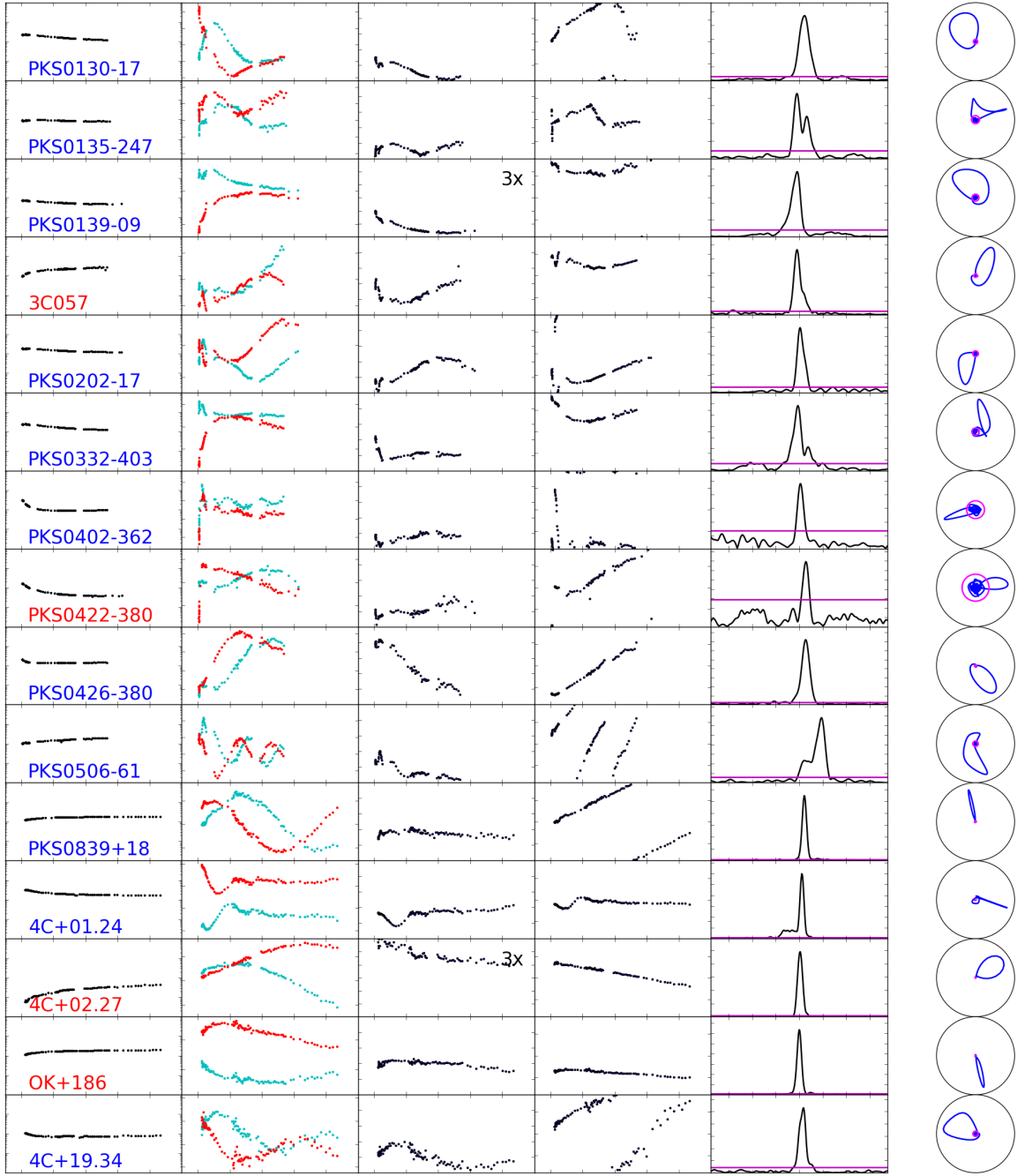}
  \caption{Measured and derived data points. Each row represents one object specified in
    Column~(i). Objects with intervener are indicated by blue and those without by red font color.
    The columns represent from left to right: (i) $I$($\lambda^2$) [0.1 Jy, 100 Jy], (ii)
    $U$($\lambda^2$) in cyan and $Q$($\lambda^2$) in red [arbitrary], (iii) $\Pi(\lambda^2)$ [0,
    10\%], (iv) $\chi(\lambda^2)$ [-90$^{\circ}$, 90$^{\circ}$], (v) $F(\phi)$ [arbitrary], (vi)
    $\mathbf{F}(\phi)$. Square brackets indicate the range of the y-axes. Column~(i) is scaled
    logarithmically and the range of Column~(iii) is enlarged to [0, 30\%] when indicated by ``3x''
    on the top right corner of the panel. The range of x-axes are [0,~0.1~m$^2$] for Columns
    (i)-(iv) and [-750~rad/m$^2$,~750~rad/m$^2$] for Column~(v). The $F(\phi)$ distributions in
    Column~(v) are shifted by the \cite{opp15} estimates of the Galactic contribution to $\phi$. The
    magenta lines in Column~(v) and (vi) represent the 5$\sigma$ noise level obtained by the sigma
    clipping algorithm. To make use of the whole circle $\chi(\phi)$, which is defined between
    -$\pi$ and $\pi$, is multiplied by 2 in Column~(vi).}
  \label{fig:data1} \end{figure*}

\begin{figure*}
  \centering
  \includegraphics[width=0.9\textwidth]{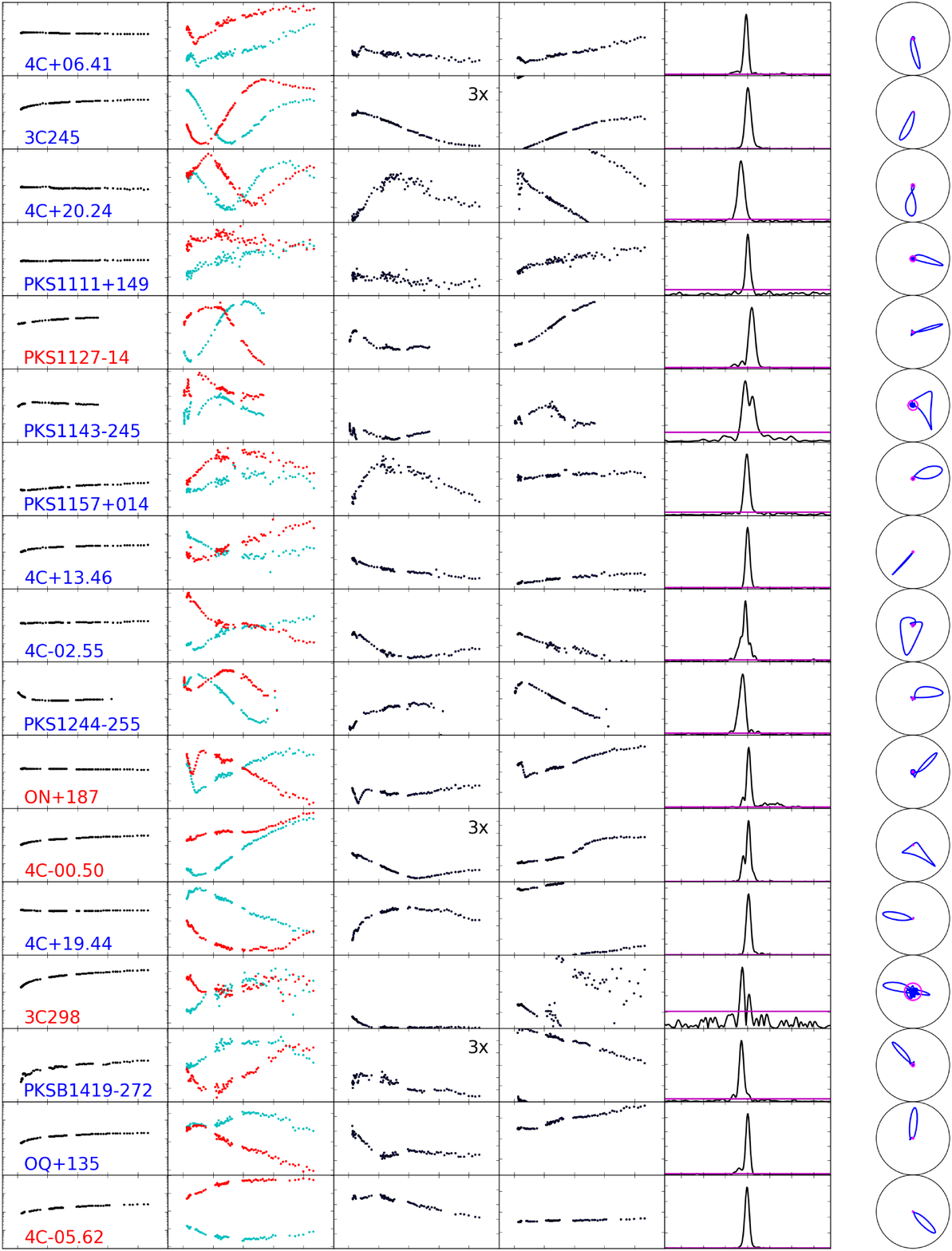}
  \caption{Compare to Figure \ref{fig:data1}.} \label{fig:data2} \end{figure*}

\begin{figure*}
  \centering
  \includegraphics[width=0.9\textwidth]{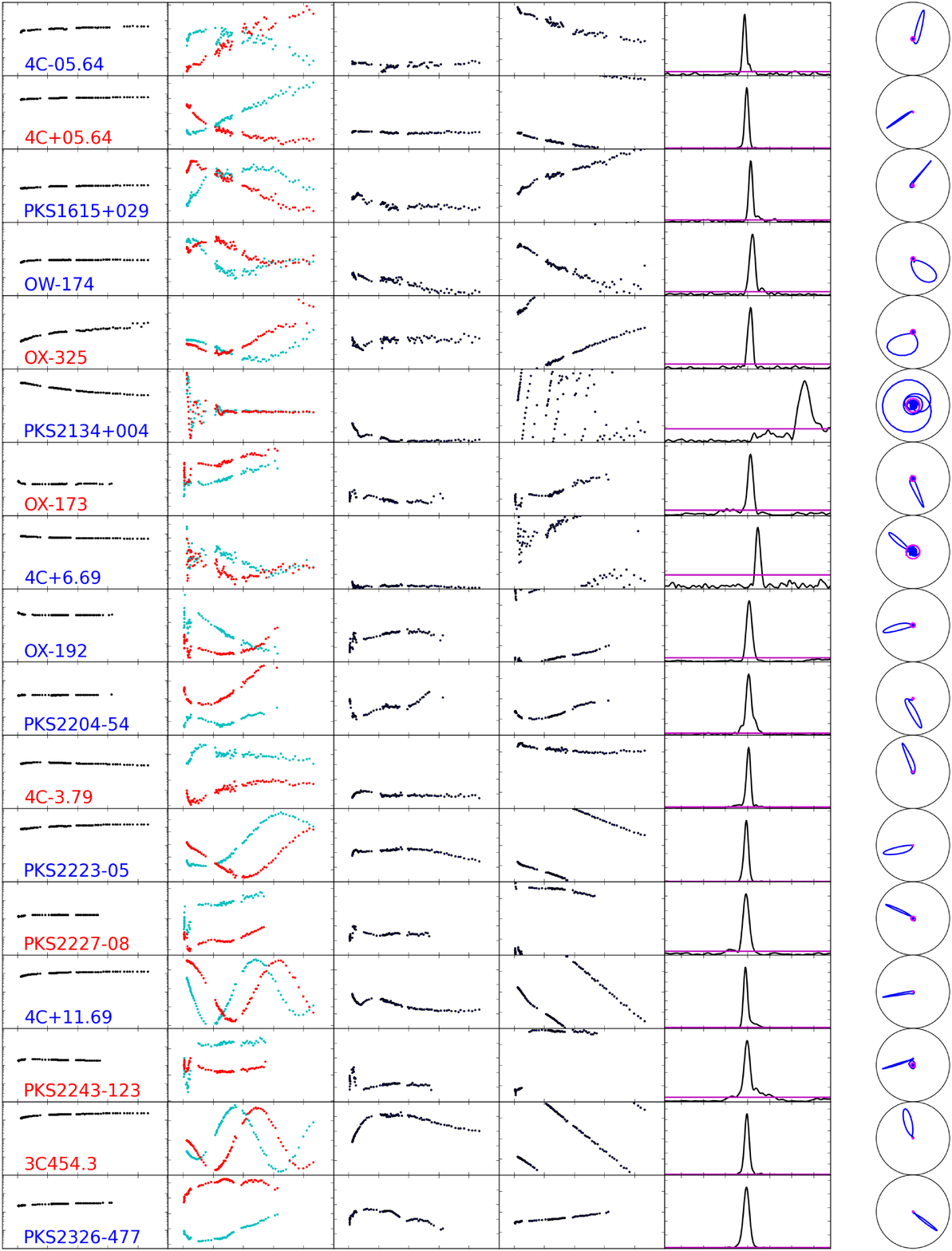}
  \caption{Compare to Figure \ref{fig:data1}.} \label{fig:data3} \end{figure*}

CASA version 4.4.0 (\cite{mul07}) has been used for all calibration steps.  3C286 served as bandpass
and flux calibrator in all three scheduling blocks. Since the objects are distributed over the sky
individual phase calibrators had to be chosen for each object. For nearby objects the same
calibrator has been used and bright and unresolved objects have been calibrated by themselves. The
chosen phase calibrators are listed in Table~\ref{tbl:phasecal}. Sometimes sources are resolved in L
and S band but not in C band. In those cases phase calibrators have been used only for L and S
band. The phase calibrators were always observed both, before and after the observation of the
target.  3C286 also serves as the polarization angle calibrator (\cite{per13}).  The unpolarized
source J1407+2827 was used to correct for polarization leakage. The residual instrumental
polarization is safely below 0.3\%.

Data that was obviously affected by radio frequency interference (RFI) have been flagged by hand. After
recalibration further flagging has been applied by running the \texttt{flagdata} command in
\texttt{rflag} mode (\cite{gre03}). Altogether around ~30\% of the data have been flagged for each
source, mostly in the L~band.

Since only the quasar itself is interesting for this work we synthesize images just in a small window of
30$''\times$30$''$ around it. The images have been cleaned according to the Cotton-Schwab algorithm
(\cite{schwab84}, \cite{cla80}, \cite{hög74}) and applying Briggs weighting with robust parameter
$R=0$.  To ensure that we do not get any flux leakage from other sources situated outside the
synthesized window it has been checked, for each source, whether there are other bright sources within
the primary beam and it has never been the case. In a few cases where the source is extended the
image sizes have been matched appropriately. In L band images are made in steps of around 16 MHz and
in S and C Band in steps of 128 MHz. Self calibration has been applied in all frames.

To make the images at different frequencies comparable, all images have been smoothed to the beam
size of the lowest resolution image, typically around 1.3$''$. To retrieve the flux density the
brightness has been integrated over an aperture covering the FWHM of the synthesized beam centered
around the maximum flux density pixel in the $I$ frame.  Subsequently $Q$ and $U$ flux densities
have been obtained by integrating over the same aperture in $Q$ and $U$. The obtained flux densities
have been converted to total flux densities assuming a two dimensional Gaussian profile of the
sources.  The retrieved data ($I$, $Q$ and $U$ parameter) as well as the derived polarization
fraction $\Pi$ and polarization angle $\chi$ are shown in the Columns~(i)-(iv) of Figures
\ref{fig:data1}-\ref{fig:data3}.

\subsection{ATCA observations, data reduction and flux density measurements}

The ATCA observations took place on 2014 May 3 (A-SB1, cf. Table~\ref{tbl:qso}) and 2013 May 14
(A-SB2), 15 (A-SB3) in configuration 6C corresponding to a maximum baseline of 6~km. The observed
frequency range was 1.1 - 3.1 GHz (16~cm band), 4.0 - 6.0 GHz (4~cm band) and 8.8 - 10 GHz (4~cm
band) with a resolution of 1 MHz (Proposal ID: C2769, PI: M. L. Bernet). Each source has been
observed with at least three cuts in the uv-plane.

The reduction package MIRIAD (\cite{sau95}) has been used to carry out the calibration. For sources
in A-SB1 PKSB1934-638 served as the bandpass calibrator in the 16~cm band and PKSB1921-293 in the
4~cm~band while for sources in A-SB2 and A-SB3 PKS0823-500 served as the bandpass
calibrator. Furthermore PKSB1934-638 is used as the flux calibrator for all sources. We observed
repeatedly PKSB2326-477 and PKSB2005-489 during A-SB1 and PKS1903-802 during A-SB2 and A-SB3 for
phase and polarization leakage calibration.  However, since the phase calibration solutions turned
out to be unsatisfying, we relied solely on self calibration for all sources. This is the reason why
some sources were discarded. The \texttt{pgflag} command has been used to auto-flag corrupted
data. Some manual flagging has been applied afterwards.

$I$, $Q$ and $U$ Images of 2$'\times$2$'$ frames around the objects are synthesized in steps of
16~MHz over the whole observed frequency band applying natural weighting. Cleaning has been carried
out with the \texttt{clean} command in \texttt{any} mode. All frames have been smoothed to the same
resolution, typically around 10$''$, and it has been ensured that they contain only one single
unresolved source. The flux density of that source has been determined by measuring the maximum
brightness of the frames. The data points of the 4~cm~band images have been binned in steps of
128~MHz to have similar data spacings as the VLA objects.

The retrieved data ($I$, $Q$ and $U$) as well as the derived polarization fraction $\Pi$ and
polarization angle $\chi$ are shown in the Columns~(i)-(iv) of Figures
\ref{fig:data1}-\ref{fig:data3}. We conclude that the structure of $I$, $Q$, $U$ is comparable
between sources observed at VLA with sources observed at ATCA ensuring the consistency of our
sample.

\begin{figure*}
  \centering
  \includegraphics[width=0.9\textwidth]{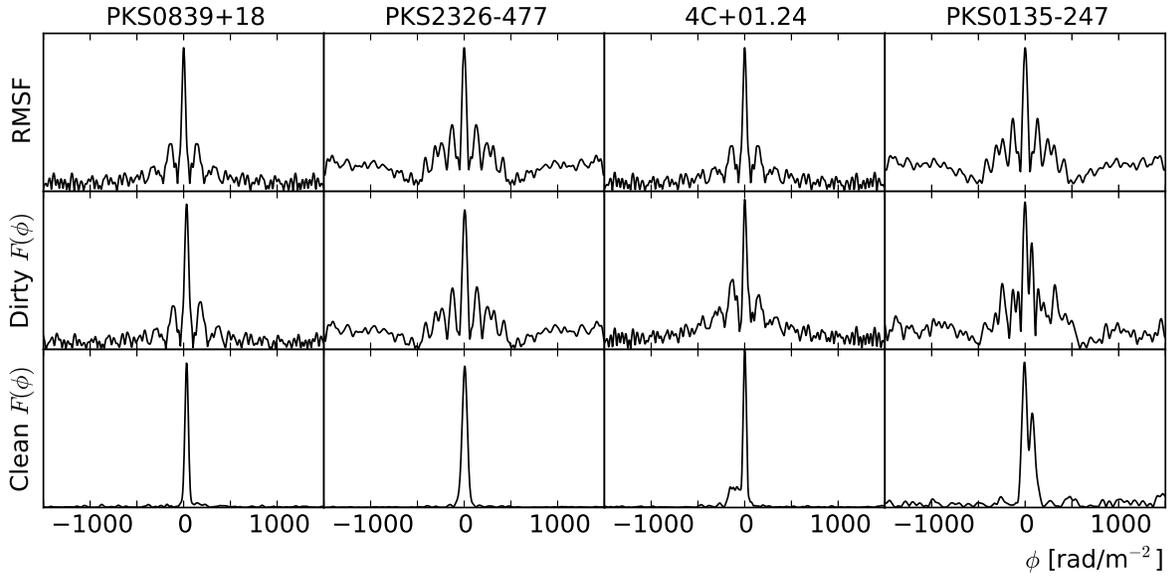}
  \caption{Effect of RM-CLEAN for four representative objects that have simple (first and second
    columns) and complex (third and fourth columns) FD distributions. The first and third column
    objects were observed with the VLA and the second and fourth column ones were observed with the
    ATCA.  The first row represents the RMSF used by RM-CLEAN, the second row the dirty FD
    distribution, i.e. before RM-CLEAN is applied, and the third row the clean FD distribution after
    RM-CLEAN. The y-axes are in arbitrary units.} \label{fig:rmsf} \end{figure*}

\subsection{Rotation Measure Synthesis} \label{sec:rmsynthesis}

RM Synthesis was carried out using the code provided by \cite{bre05}. The RM Synthesis is calculated
with uniform weighting and relative to $\lambda^2=0$.  Subsequently, the RM-CLEAN algorithm has been
applied, following \cite{hea09}. The effects of RM-CLEAN are shown in Figure~\ref{fig:rmsf} for four
objects. PKS0839+18 and PKS2326-477 are sources with simple FD distributions that were observed with
the VLA and ATCA, respectively. 4C+01.24 and PKS0135-247 represent sources with more complex FD
distributions from the VLA and ATCA, respectively. The second row shows the dirty $F(\phi)$, that is
before RM-CLEAN is applied, and the third row shows the clean $F(\phi)$ after the application of
RM-CLEAN.

The first row shows the Rotation Measure Spread Functions (RMSF) of the objects. The RMSF is the
function with which the true FD distribution is convolved due to the finite sampling of the data in
$\lambda^2$ space. After running RM-CLEAN the RMSF corresponds to a Gaussian whereby its FWHM can be
estimated as (\cite{bre05})
\begin{equation}
\delta\phi \approx \frac{2\sqrt{3}}{\Delta\lambda^2} \label{eq:rmsf}
\end{equation}
where $\Delta\lambda^2$ is the observed
frequency bandwidth in $\lambda^2$~space. For ease of comparison later on we
introduce
\begin{equation}
\sigma_\mathrm{RMSF} = \frac{\delta\phi}{2\sqrt{2\ln{2}}} \label{eq:rmsfsigma}
\end{equation}
defined as the standard deviation of the clean Gaussian RMSF. $\sigma_\mathrm{RMSF}$ corresponds to
the ``resolution" of the data in $\phi$ space. The $\sigma_\mathrm{RMSF}$ for the VLA data is around
17 rad/m$^2$ and for the ATCA data it is around 24 rad/m$^2$.

Despite some known shortcomings of the RM-CLEAN method to accurately reconstruct complex FD
distributions (\cite{sun15}) we dispense with more sophisticated but complicated methods
(e.g. \cite{fri10}, \cite{li11}, \cite{farns11}, \cite{sul12}, \cite{schn15}).  We regard the method
adopted to be sufficient for our purposes since our analysis aims primarily to separate sources with
simple FD distributions from those with complex ones. As will become clear later in the paper, our
conclusions do not depend on small components in the FD distribution.

Historically it has been conventional to use the Stokes parameter $Q$ and $U$ as input for the RM
Synthesis, according to the definition in Equation~\ref{eq:fdfourier}. Recently, however, the fractional
parameter $q=Q/I$ and $u=U/I$ has been utilized (\cite{and15}). There are good reasons to do that,
especially when one is dealing with steep spectral index sources. Since our sources, however, mostly
have flat spectra, we do not expect the differences of the two methods to be severe. Indeed we have
tried out both methods and it turns out that the conclusions we draw in this paper hold whatever
method we use. For the presentation of the results in this paper we choose to stick with the
conventional method, i.e. utilizing $Q$ and $U$, and mention the results with $q$ and $u$.

The obtained FD distributions $F(\phi)$ are shown in the panels of Column~(v) of Figure
\ref{fig:data1}-\ref{fig:data3} for all 49 sources. Although the FD distributions are synthesized
between -1500~rad/m$^2$ and 1500~rad/m$^2$ we show only the range between -750~rad/m$^2$ and
750~rad/m$^2$ since we do not detect any significant signal outside of this window. We adopt the
Galactic RM contribution estimates from \cite{opp15} and shift the overall FD distributions
accordingly.

The polar plots in the panels of Column~(vi) represent the complex FD distribution
$\mathbf{F}(\phi)$. In this, the azimuthal angle represents the initial phase $\psi$, i.e. the
polarization angle of a particular component at infinite frequency (see equation~\ref{eq:fd} above)
while the radius is a measure of the amplitude that was plotted in Column~(v). We obtain continuous
curves since both $F(\phi)$ and $\psi(\phi)$ are continuous. The initial phase $\psi$ is defined
between -$\pi/2$ and $\pi/2$ and so for plotting purposes, $\psi$ is multiplied by 2.

The FD distribution $F(\phi)$ decomposes the observed flux density in $\phi$, i.e. it describes the
amount of linearly polarized flux density which has undergone a certain amount of Faraday Rotation
due to lying at a certain Faraday Depth. Thus, for a homogeneous screen lying in front of a simple
source, the FD distribution would ideally be a delta function.  An inhomogeneous screen would result
in a broader or more complicated FD distribution because different parts of the background source
would have passed through different Faraday depths $\phi$.  In this work we are interested in
extracting information on the inhomogeneity of foreground screens and therefore the FD distribution
is what we are ultimately interested in.

However, from the FD distribution alone, we do not know if structure within $F(\phi)$ is caused by
variations within an intervening system, somewhere along the line of sight, or by having a source
which is itself Faraday thick. Sources are called Faraday thick if they have a range of $\phi$ due
to magnetized plasma intrinsic to the sources. This can occur in two ways, either through internal
Faraday dispersion or through differential Faraday Rotation (\cite{sok98}). Internal Faraday
dispersion is caused by intrinsic inhomogeneous screens. Effectively the source is composed of
different sub-components each with their own $\phi$.  Differential Faraday Rotation is caused if a
source is extended along the line of sight such that flux which is emitted from the far side of the
object undergoes more rotation than flux which is emitted from the near side.

To discriminate between intrinsic inhomogeneity effects and intervener inhomogeneity effects
$\psi(\phi)$ can be very useful. As mentioned already in Section~\ref{sec:intro} the initial phase
represents the average polarization angle of the flux density at a certain $\phi$. Thus if all the
emission has the same origin then $\psi(\phi)$ should be constant, independent of how complex the FD
distribution might be from the intervener system. If, however, $\psi(\phi)$ is not constant, then it
can be concluded that the emission at the different $\phi$ must have different spatial origins
within the source.  Correspondingly, it becomes very likely (but not absolutely proven) that the
variation of $\phi$ is also intrinsic to the source.  The case of constant $\psi(\phi)$ corresponds
to a radial locus of $\mathbf{F}$ in a polar representation, i.e. a linear feature extending out
from the central origin.  There are numerous cases of this in
Figures~\ref{fig:data1}-\ref{fig:data3} (e.g. PKS0839+18, 4C+13.46 and 4C+05.64).

To illustrate a more complex example, consider again the case of a source (with intrinsic magnetic
fields) that is extended along the line of sight. The emission coming from the far side of the source
can be produced with different polarization angle than emission which is coming from the near side
and therefore can have different initial phases. Eventually one would expect an ellipse like
structure in the polar plots of the complex FD distribution in which the initial phase
varies smoothly with the Faraday Depth. Analogously one would also get ellipse-like structures if
there are unresolved spatial polarization structures in the extended source as well as an spatially
inhomogeneous FD screen in front of it.  Examples of ellipses in
Figures~\ref{fig:data1}-\ref{fig:data3} are 4C+02.27, OW-174 and 4C-05.62.

More complex situations are also possible.  A source consisting of a number of discrete components
within the telescope beam, each with a very narrow range of $\phi$, will exhibit the corresponding
number of peaks in $F(\phi)$ (in Column~(v)) and radial spikes in $\mathbf{F}(\phi)$ (in
Column~(vi)) producing a complex amoeba-like structure in the latter.  Examples are given by
4C-00.50, PKS1143-247 and PKS0506-61. On the other hand, a simple (but extended) background source
that undergoes Faraday Rotation by a number of discrete $\phi$ within an inhomogeneous foreground
screen would consist of multiple peaks in $F(\phi)$ (in Column~(v)) but the radial spikes in
$\mathbf{F}(\phi)$ (in Column~(vi)) would be aligned. Interestingly, we see no such sources in our
sample, a point to which we will return later.

Overall, we see that there is a wide variation in complexity of $F(\phi)$ and $\mathbf{F}(\phi)$
within our sample as shown in Figures~\ref{fig:data1}-\ref{fig:data3}. However, despite this clear
variety, we also see that each object usually has a pronounced dominant Gaussian-like component in
which a large fraction of the total flux density is contained. This is in agreement with the
observations of low redshift quasars in \cite{and15}. We will refer to this component as the
``primary component'' throughout this paper. All other significant components will be referred to as
``secondary components''. Roughly two third of our sample possess significant secondary
components. Generally speaking, the structure of the FD distribution is broadly comparable between
those sources observed at VLA and those observed at ATCA, except for the different width of the
RMSF. 

\begin{turnpage}
\begin{deluxetable*}{lrrrrrrrrcccrc}
  \tablewidth{1.2\textwidth} \tabletypesize{\small} \tablecaption{Depolarization and Faraday Depth
    Parameters\label{tbl:param}} \tablehead{\multicolumn{1}{l}{Name       } &
    \colhead{$\tilde{\phi}_\mathrm{max}$$^1$}
    & \colhead{$\phi_\mathrm{max}$$^1$}
    & \colhead{DP$_\mathrm{Quart}$} & \colhead{DP$_\mathrm{Half}$} & \colhead{DP$_\mathrm{Min/Max}$}
    & \colhead{DP$'_\mathrm{Min/Max}$} & \colhead{$\sigma_\mathrm{Burn}$$^1$}
    & \colhead{$\sigma'_\mathrm{Burn}$$^1$}
    & \colhead{$\sigma_\mathrm{FD}$$^{1,2}$}
    & \colhead{$G$$^2$}
    & \colhead{$C$$^2$}
    & \colhead{Rank} & \colhead{$\sigma_\mathrm{PC}$$^{1,2}$}
  } \startdata 
PKS0130-17 & 52.2 & 43.7 & 0.902 & 0.779 & 0.976 & 0.976 & 25.17 & 25.04 & 103.5 / 52.7 & 0.069 / 0.050 & 0.111 / 0.044 & 20 & 35.7 / 33.3 \\
PKS0135-247 & -6.3 & -17.6 & -0.392 & -0.286 & 0.851 & 0.197 & 0.00 & -$^3$ & 55.4 / 116.9 & 0.054 / 0.080 & 0.366 / 0.444 & 45 & 12.0 / 11.2 \\
PKS0139-09 & -19.7 & -29.6 & 0.775 & 0.713 & 0.873 & 0.873 & 28.30 & 27.98 & 50.8 / 117.7 & 0.048 / 0.074 & 0.058 / 0.181 & 27 & 34.5 / 29.5 \\
3C057 & -0.6 & -14.6 & -0.197 & -0.430 & 0.769 & 0.297 & -$^3$ & -$^3$ & 136.0 / 136.5 & 0.050 / 0.076 & 0.140 / 0.230 & 24 & 16.8 / 17.4 \\
PKS0202-17 & 20.4 & 13.9 & -1.842 & -1.471 & 0.958 & 0.376 & -$^3$ & 10.10 & 50.7 / 99.0 & 0.042 / 0.066 & 0.078 / 0.151 & 16 & 21.2 / 23.7 \\
PKS0332-403 & -24.0 & -16.4 & 0.281 & 0.028 & 0.702 & 0.702 & 9.76 & 9.26 & 167.6 / 156.7 & 0.117 / 0.102 & 0.391 / 0.377 & 43 & 28.2 / 20.9 \\
PKS0402-362 & -0.7 & 12.4 & -1.085 & -0.289 & 0.897 & 0.399 & -$^3$ & 9.28 & 42.6 / 94.0 & 0.038 / 0.073 & 0.093 / 0.204 & 9 & 10.6 / 25.3 \\
PKS0422-380 & 39.1 & 55.3 & -1.243 & -0.661 & 0.945 & 0.542 & -$^3$ & -$^3$ & 25.7 / 89.0 & 0.026 / 0.063 & 0.000 / 0.251 & 13 & 16.2 / 9.1 \\
PKS0426-380 & 38.1 & 55.0 & 0.696 & 0.641 & 0.850 & 0.850 & 19.69 & 18.33 & 84.5 / 53.2 & 0.055 / 0.050 & 0.116 / 0.355 & 21 & 21.7 / 8.7 \\
PKS0506-61 & 113.7 & 179.9 & 0.790 & 0.609 & 0.912 & 0.912 & 24.64 & 18.24 & 80.2 / 72.4 & 0.073 / 0.068 & 0.258 / 0.030 & 48 & 31.6 / 58.0 \\
PKS0839+18 & 31.8 & 44.4 & 0.117 & 0.082 & 0.352 & 0.337 & 3.50 & 4.48 & 43.4 / 41.2 & 0.030 / 0.026 & 0.097 / 0.076 & 17 & 7.8 / 6.3 \\
4C+01.24 & 0.8 & 23.6 & -0.247 & -0.138 & 0.634 & 0.000 & -$^3$ & -$^3$ & 70.5 / 64.0 & 0.051 / 0.042 & 0.314 / 0.240 & 46 & 3.4 / 7.2 \\
4C+02.27 & -11.2 & 8.1 & 0.282 & 0.179 & 0.510 & 0.510 & 5.76 & 5.71 & 22.6 / 24.8 & 0.022 / 0.021 & 0.007 / 0.042 & 8 & 12.6 / 4.6 \\
OK+186 & -3.9 & 1.2 & 0.190 & 0.098 & 0.368 & 0.368 & 4.12 & 4.81 & 32.8 / 28.4 & 0.026 / 0.023 & 0.072 / 0.052 & 11 & 9.1 / 7.5 \\
4C+19.34 & 35.6 & 28.1 & 0.638 & 0.621 & 0.917 & 0.917 & 12.77 & 15.55 & 36.3 / 48.1 & 0.035 / 0.044 & 0.053 / 0.111 & 26 & 24.4 / 25.7 \\
4C+06.41 & 10.5 & -11.0 & 0.332 & 0.146 & 0.536 & 0.536 & 5.80 & 5.76 & 91.5 / 91.6 & 0.054 / 0.051 & 0.217 / 0.194 & 28 & 9.1 / 8.1 \\
3C245 & 23.7 & 6.8 & 0.879 & 0.715 & 0.922 & 0.922 & 14.09 & 14.33 & 37.9 / 36.7 & 0.033 / 0.029 & 0.052 / 0.034 & 15 & 19.9 / 17.4 \\
4C+20.24 & -38.8 & -56.6 & -1.467 & 0.243 & 0.973 & 0.676 & -$^3$ & 9.56 & 47.0 / 35.4 & 0.040 / 0.034 & 0.070 / 0.033 & 22 & 25.4 / 25.7 \\
PKS1111+149 & 12.5 & 5.0 & 0.381 & 0.376 & 0.808 & 0.808 & 7.68 & 7.64 & 34.9 / 34.3 & 0.028 / 0.028 & 0.090 / 0.083 & 14 & 12.0 / 12.8 \\
PKS1127-14 & 40.3 & 40.4 & 0.394 & 0.262 & 0.505 & 0.505 & 11.41 & 4.24 & 70.2 / 56.0 & 0.052 / 0.045 & 0.197 / 0.181 & 34 & 2.3 / 19.0 \\
PKS1143-245 & -0.5 & -18.0 & 0.259 & 0.425 & 0.884 & 0.884 & 15.52 & 15.19 & 82.3 / 279.3 & 0.068 / 0.095 & 0.409 / 0.375 & 47 & 15.6 / -$^4$ \\
PKS1157+014 & 2.5 & -4.8 & -0.016 & 0.219 & 0.804 & 0.774 & 4.58 & 8.46 & 62.2 / 65.2 & 0.047 / 0.047 & 0.149 / 0.173 & 19 & 17.1 / 17.2 \\
4C+13.46 & 6.1 & 2.8 & 0.580 & 0.320 & 0.683 & 0.683 & 9.46 & 9.31 & 23.7 / 92.1 & 0.021 / 0.032 & 0.027 / 0.109 & 4 & 8.0 / 8.5 \\
4C-02.55 & -18.7 & -19.6 & 0.549 & 0.515 & 0.877 & 0.877 & 28.92 & 28.96 & 113.0 / 106.8 & 0.052 / 0.051 & 0.350 / 0.151 & 41 & 12.4 / 28.7 \\
PKS1244-255 & -29.5 & -47.1 & -2.787 & -1.189 & 1.000 & 1.000 & 0.00 & 14.94 & 46.4 / 46.4 & 0.041 / 0.041 & 0.101 / 0.064 & 32 & 17.1 / 22.6 \\
ON+187 & 15.1 & 13.1 & -0.349 & -0.237 & 0.771 & 0.155 & 0.00 & -$^3$ & 133.0 / 128.3 & 0.085 / 0.081 & 0.379 / 0.367 & 44 & 6.9 / 6.4 \\
4C-00.50 & 16.5 & 10.8 & 0.615 & 0.634 & 0.864 & 0.864 & 21.83 & 21.69 & 44.7 / 48.2 & 0.036 / 0.041 & 0.257 / 0.200 & 39 & 9.6 / 17.8 \\
4C+19.44 & 8.8 & 10.2 & -0.633 & -0.093 & 0.693 & 0.280 & 0.00 & 5.29 & 37.2 / 56.6 & 0.027 / 0.036 & 0.070 / 0.097 & 18 & 10.8 / 12.5 \\
3C298 & -40.6 & -47.6 & 0.918 & 0.483 & 0.985 & 0.985 & 79.93 & 78.68 & 35.9 / 132.8 & 0.030 / 0.122 & 0.306 / 0.599 & 42 & 6.7 / 28.6 \\
PKSB1419-272 & -17.1 & -52.8 & 0.669 & 0.487 & 0.832 & 0.832 & 10.63 & 11.29 & 33.7 / 42.0 & 0.031 / 0.033 & 0.124 / 0.104 & 33 & 12.7 / 16.6 \\
OQ+135 & 12.3 & 3.8 & 0.485 & 0.109 & 0.666 & 0.666 & 7.64 & 7.29 & 69.6 / 56.5 & 0.030 / 0.040 & 0.147 / 0.248 & 31 & 6.5 / -$^4$ \\
4C-05.62 & 1.3 & -2.9 & 0.337 & 0.190 & 0.425 & 0.425 & 6.20 & 6.38 & 27.7 / 40.7 & 0.024 / 0.023 & 0.067 / 0.035 & 2 & 10.4 / 7.9 \\
4C-05.64 & -16.8 & -25.2 & 0.015 & -0.058 & 0.773 & 0.166 & -$^3$ & -$^3$ & 139.5 / 136.3 & 0.037 / 0.035 & 0.226 / 0.214 & 25 & -$^4$ / -$^4$ \\
4C+05.64 & -10.6 & -5.9 & 0.030 & 0.039 & 0.179 & 0.179 & 1.79 & 0.00 & 32.8 / 24.0 & 0.024 / 0.021 & 0.064 / 0.045 & 10 & 2.5 / 2.3 \\
PKS1615+029 & 21.9 & 30.4 & 0.416 & 0.226 & 0.585 & 0.585 & 7.44 & 6.40 & 126.3 / 123.3 & 0.050 / 0.050 & 0.235 / 0.261 & 30 & 3.4 / 6.2 \\
OW-174 & -31.5 & 44.5 & 0.858 & 0.713 & 0.982 & 0.982 & 13.39 & 13.31 & 44.2 / 43.5 & 0.035 / 0.034 & 0.110 / 0.113 & 35 & 18.6 / 17.5 \\
OX-325 & 27.0 & 25.7 & -0.249 & -0.086 & 0.461 & 0.241 & -$^3$ & -$^3$ & 27.9 / 59.3 & 0.024 / 0.040 & 0.031 / 0.167 & 7 & 13.3 / 10.1 \\
PKS2134+004 & 483.5 & 524.3 & 0.880 & 0.731 & 0.996 & 0.996 & 42.02 & 39.44 & 207.4 / 197.5 & 0.182 / 0.136 & 0.396 / 0.406 & 49 & 63.3 / 47.2 \\
OX-173 & 18.3 & 28.4 & 0.278 & 0.208 & 0.512 & 0.512 & 7.39 & 3.88 & 87.2 / 30.4 & 0.054 / 0.030 & 0.287 / 0.022 & 37 & 8.2 / 14.4 \\
4C+6.69 & 29.8 & 96.2 & 0.527 & 0.225 & 0.932 & 0.932 & 7.13 & 6.50 & 610.2 / 547.3 & 0.036 / 0.038 & 0.229 / 0.193 & 5 & 12.9 / 15.0 \\
OX-192 & 11.9 & 17.1 & -0.408 & -0.230 & 0.704 & 0.340 & -$^3$ & 7.30 & 29.0 / 30.3 & 0.028 / 0.029 & 0.019 / 0.020 & 6 & 10.5 / 13.2 \\
PKS2204-54 & 12.2 & 13.7 & -0.727 & -0.103 & 0.635 & 0.127 & -$^3$ & -$^3$ & 102.9 / 49.7 & 0.060 / 0.042 & 0.206 / 0.131 & 23 & 15.4 / 8.9 \\
4C-3.79 & -5.5 & 10.4 & 0.196 & 0.035 & 0.409 & 0.409 & 3.56 & 2.39 & 63.1 / 48.1 & 0.046 / 0.034 & 0.197 / 0.143 & 38 & 8.9 / 5.5 \\
PKS2223-05 & -23.5 & -10.3 & 0.178 & 0.056 & 0.410 & 0.410 & 3.96 & 6.45 & 25.5 / 25.0 & 0.022 / 0.022 & 0.028 / 0.038 & 1 & 10.6 / 7.5 \\
PKS2227-08 & -11.0 & -11.7 & -0.025 & -0.053 & 0.511 & 0.349 & 1.80 & -$^3$ & 62.0 / 58.4 & 0.048 / 0.046 & 0.133 / 0.139 & 29 & 9.4 / 6.9 \\
4C+11.69 & -52.4 & -16.5 & 0.361 & 0.308 & 0.521 & 0.521 & 7.26 & 7.13 & 53.9 / 81.5 & 0.034 / 0.055 & 0.155 / 0.211 & 36 & 8.1 / 8.9 \\
PKS2243-123 & -15.9 & -2.5 & 0.273 & 0.138 & 0.654 & 0.654 & 9.07 & 6.54 & 98.8 / 61.6 & 0.086 / 0.044 & 0.344 / 0.064 & 40 & 12.8 / 22.4 \\
3C454.3 & -51.7 & -4.4 & -0.091 & 0.114 & 0.478 & 0.301 & 1.10 & 5.04 & 44.1 / 42.8 & 0.030 / 0.029 & 0.079 / 0.077 & 12 & 12.6 / 10.7 \\
PKS2326-477 & 5.9 & -8.7 & 0.180 & 0.226 & 0.521 & 0.521 & 6.63 & 9.44 & 50.3 / 96.2 & 0.038 / 0.083 & 0.066 / 0.353 & 3 & 15.4 / 9.3
\enddata
\tablenotetext{1}{in units of [rad/m$^2$]. The $\tilde{\phi}_\mathrm{max}$ parameter is before and
  the $\phi_\mathrm{max}$ parameter after Galactic foreground subtraction.}  \tablenotetext{2}{The second numbers
  are the values when fractional $u$ and $q$ is utilized for RM Synthesis
  (cf. Section~\ref{sec:rmsynthesis}).}  \tablenotetext{3}{The fit has failed to converge or
  unphysical values has been obtained.}  \tablenotetext{4}{$\sigma_\mathrm{RMSF}$ has been larger
  than $\sigma_\mathrm{fit}$. \newline}
\end{deluxetable*}
\end{turnpage}

\section{Parameters}\label{sec:parameters}

In this section we define all the parameters which will be used in Section~\ref{sec:results} to
quantify structures in polarization and in the FD distribution. Note that these parameters were
defined blind to the presence of interveners, in the sense that we defined and measured them before
linking the radio data with the optical spectra. The sole exception was $\sigma_{PC}$ for reasons
which will become clear later. The parameter values of all objects are listed in
Table~\ref{tbl:param}.

\subsection{Rotation Measure and $\phi_\mathrm{max}$}\label{sec:pararm}

As is clear in Figures~\ref{fig:data1}-\ref{fig:data3}, essentially every object (with the possible
exception only of PKS2134+004, see below) has a FD distribution that is dominated by a single
pronounced primary component at a well-defined $\phi_\mathrm{max}$. We define
$\tilde{\phi}_{\mathrm{max}}$ as the peak position as observed, and $\phi_\mathrm{max}$ as the peak
position after shifting the FD distribution by a uniform $\Delta \phi$ corresponding to an estimate
of the Galactic foreground according to \cite{opp15}.

It is therefore of interest to compare the $\phi_\mathrm{max}$ for each source with the
RM$_\mathrm{Kron}$ of \cite{kro08} used in \cite{bern08}. These were based on polarization angle
measurements at just a few distinct frequencies. As discussed in \cite{bern12}, there are some
caveats to this traditional way of measuring RM, especially if the sources are significantly
depolarized. These problems are largely circumvented in RM Synthesis.

\begin{figure}
    \centering\includegraphics[width=0.45\textwidth]{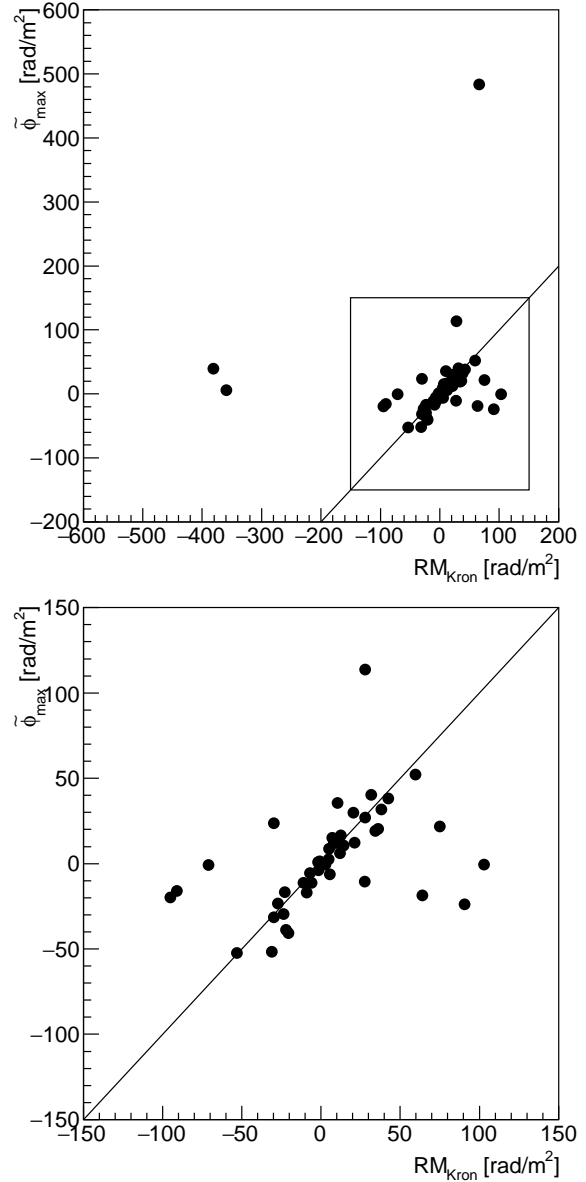}
    \caption{Comparison between RM$_\mathrm{Kron}$ from \cite{kro08} and the Faraday Depth
      $\tilde{\phi}_\mathrm{max}$ at which the FD distribution before Galactic RM subtraction has
      its maximum for all objects. The lower panel is a zoomed in depiction of the boxed region in
      the upper panel.}
    \label{fig:KrnCor}
\end{figure}

In Figure~\ref{fig:KrnCor} we compare the primary component peak $\tilde{\phi}_{\mathrm{max}}$ of
the FD distribution, before any Galactic RM correction, to the RM$_{\mathrm{Kron}}$ which were used
(similarly uncorrected) in \cite{bern08}. For about three quarters of the objects, the agreement
between these quite independent measurements is very good with a random dispersion of roughly
8~rad/m$^2$. For 12 objects there is a significant discrepancy.  Many of the objects with strong
$|\mathrm{RM_{Kron}}|>50$~rad/m$^2$ are found to have rather smaller
$|\tilde{\phi}_{\mathrm{max}}|$.  There are three objects with very large discrepancies.  The case
of PKS2134+004 is interesting in that the dominant peak is clearly at very high $\tilde{\phi}$. This
source shows extraordinarily complex FD structures (Figure~\ref{fig:data3}, Row 6).  The position
angle $\chi(\lambda^2)$ plot for this source emphasizes the difficulty of measuring an RM, for such
a source, from sparsely sampled data. In the main, however, this comparison is reassuring that the
original RM measurements used in \cite{bern08} were meaningful.

\subsection{Depolarization}\label{sec:dfdep}
  
Many years ago, \cite{bur66} calculated the relation between depolarization and a dispersion in the
Faraday Depth to be of the form
\begin{equation}
  \Pi(\lambda^2) = \Pi_0 \exp(-2(\sigma_{\mathrm{Burn}}\lambda^2)^2) \label{eq:burn}
\end{equation}
where $\Pi(\lambda^2)$ is the polarization fraction as a function of $\lambda^2$, $\Pi_0$ is the
intrinsic polarization fraction before the Faraday depolarization, which is assumed to be constant
with $\lambda$, and $\sigma_{\mathrm{Burn}}$ is the standard deviation of a Gaussian FD
distribution. Subsequently, various modifications to this simple model have been proposed to better
fit observations (\cite{tri91}, \cite{ros08}, \cite{bern12}).

All these models produce a monotonic decrease of $\Pi(\lambda^2)$ with $\lambda$.  The measured
polarization of many of our sources, however, clearly shows a more complex structure in
$\Pi(\lambda^2)$. We will explore possible reasons for this in Section~\ref{sec:toymodel}. An
extensive overview of depolarization due to Faraday screens can be found in the appendix of
\cite{far14}.

We want to test the suggestion in \cite{bern12} that intervening MgII absorption along the line of
sight also produces detectable depolarization effects. We therefore need to quantify the
depolarization in the radio spectra. However, the variety and complexity of the polarization spectra
(Column~(iii) in Figure~\ref{fig:data1}-\ref{fig:data3}) makes it impossible to represent the
depolarization with a single parameter. We therefore defined several parameters.  Although some of
these may appear somewhat \textit{ad hoc}, this was done blind to the presence of interveners.

We first define DP$_\mathrm{Quart}$ and DP$_\mathrm{Half}$
as
\begin{equation}
\mathrm{DP_{Quart,Half}}=1-\frac{\Pi_{2}}{\Pi_{1}}
\end{equation}
where $\Pi_1$ is the median $\Pi$ in the first quarter of the observed $\lambda^2$ band, i.e. in the
short wavelength end, and $\Pi_2$ the median $\Pi$ in the last quarter for DP$_\mathrm{Quart}$,
i.e. in the long wavelength end. For DP$_\mathrm{Half}$, $\Pi_1$ is the median $\Pi$ in the first
half of the observed $\lambda^2$ band and $\Pi_2$ the median $\Pi$ in the other half. By definition
DP$_\mathrm{Quart}$ and DP$_\mathrm{Half}$ yield negative values for re-polarizing sources.

We also define DP$_\mathrm{Min/Max}$ as
\begin{equation}
\mathrm{DP_{Min/Max}}=1-\frac{\Pi_\mathrm{min}}{\Pi_\mathrm{max}}
\end{equation}
where $\Pi_\mathrm{min}$ and $\Pi_\mathrm{max}$ are the minimum and maximum observed $\Pi$,
respectively wherever they occur in the observed $\lambda$ range. We also define
$\mathrm{DP'}_\mathrm{Min/Max}$, which is similar to DP$_\mathrm{Min/Max}$ except that
$\Pi_\mathrm{min}$ is the minimum polarization at a longer wavelength than the maximum
$\Pi_\mathrm{max}$. DP$_\mathrm{Quart}$, DP$_\mathrm{Half}$, DP$_\mathrm{Min/Max}$ and
$\mathrm{DP'}_\mathrm{Min/Max}$ are completely phenomenological parameters.

Lastly we introduce $\sigma_\mathrm{Burn}$ and $\sigma'_\mathrm{Burn}$ as two further depolarization
parameters. The $\sigma_\mathrm{Burn}$ parameter is defined as in equation~\ref{eq:burn} and is
obtained by fitting this relation to $\Pi(\lambda^2)$ over the whole observed wavelength range while
the $\sigma'_\mathrm{Burn}$ parameter is obtained by only fitting $\Pi(\lambda^2)$ at wavelengths
longer than $\Pi_\mathrm{max}$. The latter is motivated by the conjecture that depolarization might
be mainly caused by a simple dispersion in Faraday Depth while other more complex behavior of the
polarization structure might be caused by other processes.  The fits are carried out by using the
method of least squares.  Although $\sigma_\mathrm{Burn}$ and $\sigma'_\mathrm{Burn}$ might seem
like the most appropriate parameters if the FD distribution was Gaussian, they might be less
appropriate for the more complex polarization curves.  The values of the six depolarization
parameters for each object are listed in Table~\ref{tbl:param}.

\subsection{FD distribution}\label{sec:deffddist}

As with depolarization, it is not trivial to identify the most suitable parameters to characterize
the structure in the FD distributions due to the variety of the observed distributions (see
Figures~\ref{fig:data1}-\ref{fig:data3}, Column~(v)).  As in the previous section, we approached
this problem by defining several parameters. To preserve objectivity, we again defined all the
parameters in this section blind to the presence of interveners in the sources.

An important point is that, because RM Synthesis behaves like a Fourier Transform, random ``white"
noise in the input Q and U spectra will produce a non-zero contribution to the FD distribution
$F(\phi)$ that should be independent of $\phi$ and extends out to the $\phi$ corresponding to the
spectral resolution of the input data, in our case $\pm 1500$~rad/m$^2$.  In practice this noise
contribution to the FD distribution will itself have structure due to noise in the data. To avoid
including this spurious contribution into our characterization of the structure in the FD
distribution, the following thresholding scheme was implemented.  We first apply an iterative
sigma-clipping algorithm (clipping at 3$\sigma$) to $F(\phi)$ over the whole $\pm 1500$~rad/m$^2$
range to estimate the background mean and standard deviation of $F(\phi)$ that arises from this
noise.  We then identify real signals in $F(\phi)$ as being those regions which exceed this mean
plus 5$\sigma$, in the sense that we can be reasonably confident that $F(\phi)$ in these regions is
not coming from noise. These are the parts of $\phi$ space that we want to use to characterize the
FD distribution.  This threshold is indicated by the magenta lines in Column~(v) of
Figures~\ref{fig:data1}-\ref{fig:data3}.  For each such region above the threshold, we then extend
in each direction out to the $\phi$ at which $F(\phi)$ first crosses the mean background level.  All
other regions are set to zero, producing a thresholded $\hat{F}(\phi)$ function that can then be
parameterized as described below.

\begin{figure*}
  \centering\includegraphics[width=\textwidth]{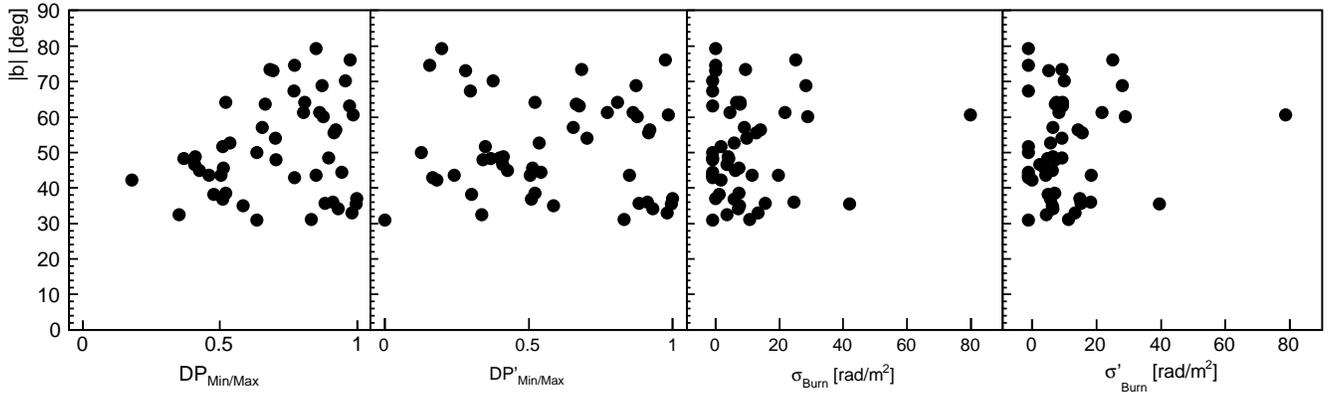}
  \caption{Comparison of absolute Galactic latitude $|b|$ to the depolarization parameters
    DP$_\mathrm{Min/Max}$ (left), DP$'_\mathrm{Min/Max}$ (middle left), $\sigma_\mathrm{Burn}$
    (middle right) and $\sigma'_\mathrm{Burn}$ (right), respectively. Objects at high $|b|$ seem to
    have enhanced DP$_\mathrm{Min/Max}$. There is no obvious trend for DP$'_\mathrm{Min/Max}$,
    $\sigma_\mathrm{Burn}$ and $\sigma'_\mathrm{Burn}$ with $|b|$.}
  \label{fig:galdep}\end{figure*}

The first parameter is simply the second moment of the modified FD distribution $\hat{F}(\phi)$
\begin{equation}
  \sigma_{\mathrm{FD}}^2 = \frac{\int{\hat{F}(\phi)(\phi-\mu_{\mathrm{FD}})^2 d\phi}}{\int{\hat{F}(\phi)d\phi}}
\end{equation}
where $\mu_\mathrm{FD}$ is the usual first-moment of the FD distribution.  Because RM Synthesis
yields a discrete FD distribution at equidistant $\phi_i$, this integral is conveniently done as a
sum over the non-zero data points. This parameter has also been used in \cite{and15}.

The second parameter is the reversed Gini coefficient $G$. The reversed Gini coefficient is simply
defined (for our equidistant data points) as
\begin{equation}
  G=1-\frac{1}{2\mu_\mathrm{FD}n(n-1)}\sum_{i}^n{\sum_{j}^n{|\hat{F}(\phi_i)-\hat{F}(\phi_j)|}} \;.
\end{equation}
where the sum is carried out over the full range of $\phi$ so that $n$ is the same for all objects.
This means that objects can be directly compared to each other, even though the absolute value of
$G$ depends on the range of $\phi$ (i.e. on $n$).  The Gini coefficient is generally used to measure
the concentration of a distribution, e.g. in economics to measure the concentration of wealth within
a population (\cite{gin12}) or in astronomy to measure the luminosity concentration in a galaxy
(\cite{abr03}). $G$ is bounded between 0 and 1 whereby 1 means equal distribution and 0 means total
concentration. We choose the reversed Gini coefficient because we want to maintain the convention
that non-concentrated FD distributions yield large parameters.  In practice, since many sources are
dominated by a single component, $G$ correlates quite well with $\sigma_\mathrm{FD}$.

We also introduce a ``coverage parameter" $C$. This parameter is motivated by the suggestion that
interveners may only partially cover the source (\cite{ros08}, \cite{man09}, \cite{bern12}). In this
scenario, we could interpret that the primary component in the FD distribution represents the flux
which is not covered by the intervening screen. We model the primary
component by fitting a Gaussian to it, using the method of least squares, and subtracting that fit
from $\hat{F}(\phi)$ distribution to yield $\hat{F}_\mathrm{cov}(\phi)$. Our coverage parameter is
then determined as
\begin{equation}
C=\frac{\int{\hat{F}_\mathrm{cov}(\phi)d\phi}}{\int{\hat{F}(\phi)d\phi}}.
\end{equation}
We obtain values of $C$ up to around 40\% in our sample. 

With the parameters above we have attempted to describe the variety of complexity that we clearly
can see. As a final subjective classification we ranked the objects in order of perceived
complexity. Two of us (KSK and SJL) independently ranked the FD distributions.  Despite the obvious
subjectivity, these rankings agreed well, except for the roughly one third of the sources which have
strikingly simple and therefore very similar FD distributions. However, the Kolmogorov Smirnov (KS)
test which we will be using throughout this paper, is relatively robust against permutations within
only the low ranked fraction of the objects. This method is to be regarded mostly as a test whether
the quantitative parameters we used do indeed represent the perceived complexity. This subjective
classification was again done blind to the presence of MgII absorption.

For reasons which will become clearer later (cf. Section~\ref{sec:toymodel}) we introduce one more
parameter $\sigma_\mathrm{PC}$, which has been defined \textit{a posteriori}. The
$\sigma_\mathrm{PC}$ parameter is obtained by fitting a Gaussian to the primary component. The range
for this fit is where the FD distribution $F(\phi)$ stops declining on either side of the primary
component.

To correct for the intrinsic spread function (RMSF) in $\phi$ that arises from the finite wavelength
band-width of the radio data, which is different for the VLA and ATCA observations, we subtract in
quadrature the predicted RMSF from the fitted $\sigma_\mathrm{fit}$ as follows:
\begin{equation}
\sigma_\mathrm{PC} = \sqrt{\sigma_\mathrm{fit}^2-\sigma_\mathrm{RMSF}^2}
\end{equation}
where $\sigma_\mathrm{RMSF}$ is defined in equation \ref{eq:rmsfsigma}. The obtained values of all
five FD distribution parameters for each object are listed in Table~\ref{tbl:param}.

\section{Effects with Galactic latitude}\label{sec:gal}

As discussed earlier, Faraday Rotation can occur anywhere along the line of sight, including within
the Galaxy, and it is therefore of interest to see if any of the parameters measured in the previous
section correlate with the Galactic latitude $b$. We will discuss in detail the depolarization
parameters DP$_\mathrm{Min/Max}$, DP$'_\mathrm{Min/Max}$, $\sigma_\mathrm{Burn}$ and
$\sigma'_\mathrm{Burn}$ and $\phi_\mathrm{max}$. These are most meaningful because they should be
independent on the instrument (VLA or ATCA) other than the FD distribution parameters such as
$\sigma_\mathrm{FD}$ and $\sigma_\mathrm{PC}$. However, we have checked these two parameters as well
and have not found any obvious trends with $b$.

Figure~\ref{fig:galdep} shows the four depolarization parameters plotted against $|b|$.
Reassuringly, there is very little correlation in these plots except for an apparent absence of
sources in the left-most panel with low DP$_\mathrm{Min/Max}$ at high Galactic latitudes.  Recall
that low DP$_\mathrm{Min/Max}$ corresponds to low depolarization, or more specifically a flat
polarization in wavelength.  We have no explanation as to why sources at high Galactic latitude
would have more complex polarization. We note that the effect is \textit{not} seen in the similar
DP$'_\mathrm{Min/Max}$ parameter (in the adjacent panel). We are confident that this possible effect
in no way drives the results which will be quoted later(Section~\ref{sec:depol}). In fact, if we
exclude all the high Galactic latitude sources with $|b| > 55^\circ$ from the analysis, the formal
significance of the association between DP$_\mathrm{Min/Max}$ and the presence of intervening MgII
absorption in the remaining sources actually increases.

\begin{figure}
  \centering\includegraphics[width=0.45\textwidth]{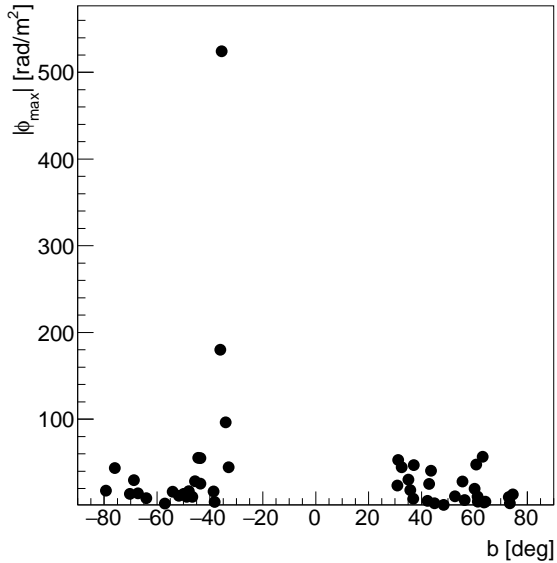}
  \caption{Comparison of Galactic latitude $b$ to the maximum peak position of the FD
      distribution $\phi_\mathrm{max}$ after Galactic RM subtraction. There is no overall trend of
      $\phi_\mathrm{max}$ with $b$ except for the two or three outlying objects with excessively
      large $\phi_\mathrm{max}$ which are all at around $|b|\approx35^\circ$.}
  \label{fig:galfrm}
\end{figure}

Turning to $\phi_\mathrm{max}$, Figure~\ref{fig:galfrm} shows this parameter, after our correction
for the Galactic RM, according to the \cite{opp15} estimates, as a function of $b$.  It is clear
that the three highest values of $|\phi_\mathrm{max}| > 80$~rad/m$^2$ are all found within a small
range of $b$ around $b \approx -35$$^\circ$,
and are therefore suspect. They are spread in Galactic longitude (at 55$^\circ$,
at 64$^\circ$
and at 277$^\circ$)
and are in regions of not very high Galactic RM in the maps provided by \cite{opp15}. Curiously,
however, two of the three are at local maxima. Regardless of the origin, these three values are
clearly suspect.  These three sources also have intervening MgII absorption systems. Therefore they
are excluded from the $\phi_\mathrm{max}$ analysis in Section~\ref{sec:rm}.

\section{Results}\label{sec:results}

\subsection{Rotation Measure and $\phi_\mathrm{max}$}\label{sec:rm}

\begin{figure}
  \centering\includegraphics[width=0.45\textwidth]{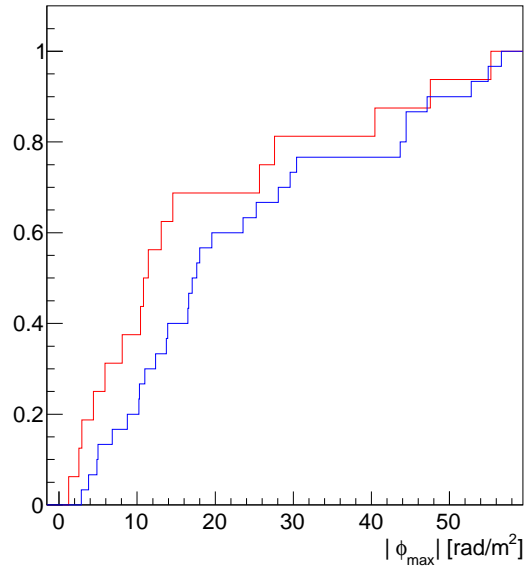}
  \caption{Empirical distribution function (EDF) of $|\phi_\mathrm{max}|$ for objects without (red)
    and with (blue) interveners. The one-tailed KS test yields $p=17\%$.}
  \label{fig:KrnCum}
\end{figure}

In \cite{bern08}, it was claimed that there was an association between higher $|$RM$|$ values and
the presence of strong (rest equivalent width above 0.3~\AA) MgII absorption in the optical
spectrum. This was based on a KS test on the distribution of $|$RM$|$ for sources with and without
such absorption.  Here we reproduce this analysis but now using our new $|\phi_{\mathrm{max}}|$
instead of $|\mathrm{RM_{Kron}}|$. $|\phi_{\mathrm{max}}|$ is the modulus of the Faraday Depth at
which the FD distribution peaks, shifted by the \cite{opp15} estimates of the Galactic contribution
to $\phi$. We use the one-tailed Kolmogorov-Smirnov to test the hypothesis that objects with MgII
absorption along the lines of sight have enhanced $|\phi_\mathrm{max}|$ (i.e. our null hypothesis is
that there is no such association or that clean lines of sight have enhanced
$|\phi_{\mathrm{max}}|$). Recall that objects with $\phi_\mathrm{max} > 80$~rad/m$^2$,
i.e. PKS0506-61, PKS2134+004 and 4C+6.69, are excluded from the analysis since we believe they are
strongly affected by Galactic effects (see Section~\ref{sec:gal}).

Unlike \cite{bern08}, we adopt a weaker absorption cut at $W_0=0.1$~\AA~and get $p=17\%$,
represented in Figure~\ref{fig:KrnCum}. We will use this cut throughout the remainder of the paper
because this cut turns out to be more significant later on when we analyze the distribution
structure of Faraday Depth indicating that also weak absorbers affect the FD distribution.  However,
also using the same equivalent width cut at 0.3~\AA~in MgII absorption, we do not see a signal and
obtain a p-value of 23\%. Using fractional $q$ and $u$ for RM Synthesis as discussed in
Section~\ref{sec:rmsynthesis} makes effectively no difference to the significance level of this test
for $|\phi_\mathrm{max}|$.

With our new data, the correlation of MgII with $|\phi_\mathrm{max}|$ is less significant than found
with RM in \cite{bern08}. They found a significance level of 92.2\% in a two-tailed KS~test, without
even correcting for Galactic RM.

This difference may be partly due to the fact that we are only analyzing a subset of their
sample. On the other hand, it could also reflect differences between the use of $\phi_\mathrm{max}$
derived from RM Synthesis (as here) and the use of single RM values that are derived from individual
sparsely-sampled polarization data (as in \cite{bern08}). Indeed, when we do the KS~test with
RM$_\mathrm{Kron}$ for our set of sample (excluding PKS0506-61, PKS2134+004 and 4C+6.69 as with our
$\phi_\mathrm{max}$ analysis) we obtain a somewhat stronger result with $p=13\%$. This stronger
significance is driven mostly by the objects with $|\mathrm{RM_{Kron}}|>50$~rad/m$^2$ which is where
we see discrepancies between $\phi_\mathrm{max}$ and RM$_\mathrm{Kron}$, as mentioned in
Section~\ref{sec:pararm}. This difference between $\phi_\mathrm{max}$ and RM$_\mathrm{Kron}$ will be
explored further in a future paper.

Prompted by the statistical editor, we also have performed the Anderson-Darling (AD) 2-sample test
for all our KS test results (here and later). It turns out that the AD test results are always
slightly more significant but altogether consistent with the KS test results. Therefore we remain
with the KS test throughout this paper.

It should be noted that the sources with intervening absorption systems are generally at slightly
higher redshifts (difference of around 0.2 in their respective medians) and so the simple dependence
of $\phi$ on wavelength would produce, all other things being equal, a lower observed dispersion for
the more distant systems, i.e. opposite to what is seen in Figure~\ref{fig:KrnCum}.

\begin{figure}
  \centering\includegraphics[width=0.45\textwidth]{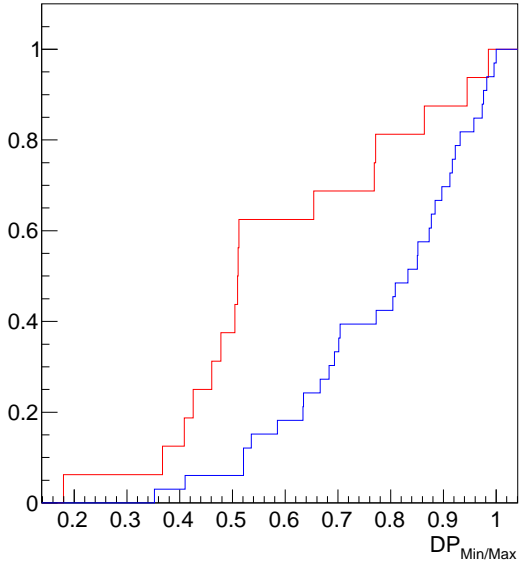}
  \caption{EDF of the depolarization parameter DP$_\mathrm{Min/Max}$ for objects without (red) and
    with (blue) interveners. The one-tailed KS test yields $p=0.1\%$.}
  \label{fig:DepCum}
\end{figure}

\begin{figure*}
  \centering\includegraphics[width=\textwidth]{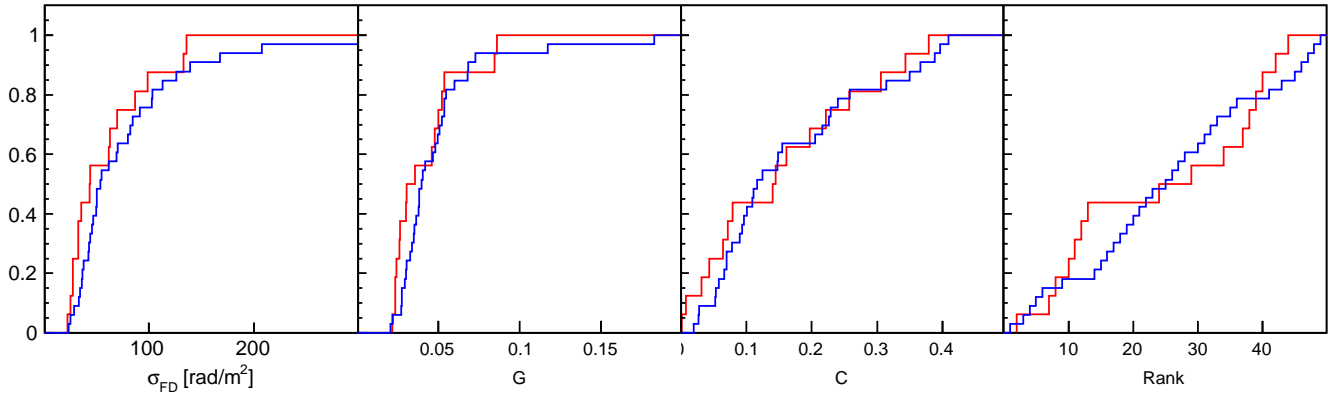}
  \caption{EDF of second moment parameter $\sigma_\mathrm{FD}$ (left), Gini coefficient $G$ (middle
    left), Coverage parameter $C$ (middle right) and subjective ranking (right) for objects without
    (red) and with (blue) interveners. The one-tailed KS test yields $p=32\%$, $p=17\%$, $p=49\%$
    and $p=24\%$ respectively.}
  \label{fig:fdist}\end{figure*}

\subsection{Depolarization}\label{sec:depol}

We now look for correlations between the depolarization parameters and the presence of MgII
absorption lines along the line of sight.  As before, we apply the one-tailed KS test with the null
hypothesis that sources with MgII absorption along the lines of sight are drawn from the same
distribution as objects with clean lines of sight or that objects with clean lines of sight are more
strongly depolarized.

The strongest signal is achieved with DP$_\mathrm{Min/Max}$ for which the KS test yields a p-value
of only $p=0.1\%$. Figure~\ref{fig:DepCum} shows the distribution functions of
DP$_\mathrm{Min/Max}$. As already mentioned in Section~\ref{sec:gal}, if objects at high Galactic
latitude $|b|>55^\circ$ are excluded the signal increases to a nominal $p=0.01\%$.

Also DP$'_\mathrm{Min/Max}$ yields a strong result with $p=1.3\%$.
$\sigma'_\mathrm{Burn}$ is not as robust, due to the complex behavior of $\Pi(\lambda^2)$ in many
sources, and for 10 objects the fit either fails to converge or yields negative
$\sigma_\mathrm{Burn}$ which is unphysical. Once these objects are excluded from the KS test,
however, $\sigma'_\mathrm{Burn}$ yields $p=0.5\%$.

To evaluate the real significance of the obtained p-values (here and elsewhere in the paper) we have
randomly re-scrambled objects with and without MgII absorption. The re-scrambling has been realized
by randomly drawing without replacement 16 objects and declaring them artificially to be objects
with clean lines of sight while the remaining 33 objects are declared as objects with
interveners. Subsequently the KS~test has been carried out to obtain p-values. The distribution of
p-values (after 10,000 such draws) showed that the p-value estimates are conservative in the sense
that p-values smaller than $p=0.1\%$ happen in 0.09\%, $p=1.3\%$ in 0.6\%, and $p=0.5\%$ in 0.2\% of
the realizations. This check shows that the p-values of the KS~tests are conservative.

Somewhat weaker but still significant results are also achieved for
DP$_\mathrm{Half}$ and $\sigma_\mathrm{Burn}$ with $p=3.9\%$ and $p=9.0\%$. As with
$\sigma'_\mathrm{Burn}$ the fit of $\sigma_\mathrm{Burn}$ fails for 9 objects and those are excluded
from the KS test for $\sigma_\mathrm{Burn}$.  For DP$_\mathrm{Quart}$ we see little or no effect
with $p=15\%$. 

It is clear that we obtain rather different p-values with the six depolarization parameters and we
will discuss this further below in Section~\ref{sec:toymodel}. Nonetheless, altogether our results
demonstrate that there is a clear and highly significant correlation between the presence of
intervening MgII absorption systems and depolarization.

\cite{far15} compared fractional polarization spectral indices to the presence of MgII absorption
with 41 objects and could not see any correlation between polarization structure and presence of
interveners. However, recall that our sources are initially selected to be compact while the sample
in \cite{far15} reduces to 15 when only flat spectrum sources are selected. Furthermore in
\cite{far15} only a handful of data points along the spectrum has been available which could lead to
imperfect description of the polarization structure given the complexity we can find for some sources.
\newline

\subsection{Structure in the FD distribution}\label{sec:nullresult}

The previous section demonstrated that depolarization in the radio spectra is clearly statistically
associated with the presence of intervening MgII absorption along the line of sight.  Since,
generically, depolarization reflects the presence of different $\phi$ coming from different parts of
the source, we would then expect to see correlations between the presence of interveners and various
parameters that capture the range of $\phi$ in a given source.

Armed with the quantitative parameters defined in Section~\ref{sec:deffddist}, we then
applied the KS tests.  As before, we adopt the null hypothesis that the parameters are equally
distributed between objects with and without intervening absorption systems or are enhanced for
objects without interveners.

We find no significant cause to reject the null hypothesis, obtaining $p=17\%$ for
$\sigma_\mathrm{FD}$, $p=12\%$ for $G$, $p=33\%$ for $C$ and $p=24\%$ for the subjective
ranking. Figure \ref{fig:fdist} shows the distributions of the four parameters and it is
fairly clear that there is no significant correlation between them and the presence of intervening
absorbers. The results remain insignificant when the fractional $q$ and $u$ are used for RM
Synthesis with $p=15\%$ for $\sigma_\mathrm{FD}$, $p=32\%$ for $G$ and $55\%$ for $C$.

These null results were surprising: We clearly see the connection between interveners and
depolarization in Section~\ref{sec:depol}, but cannot then associate the complexity in the FD
distribution, which we would have expected was the cause of the depolarization, with the presence of
interveners.  Furthermore, we actually see very little correlation between our parameterizations of
FD structure and depolarization.

\begin{figure*}
\centering\includegraphics[width=0.95\textwidth]{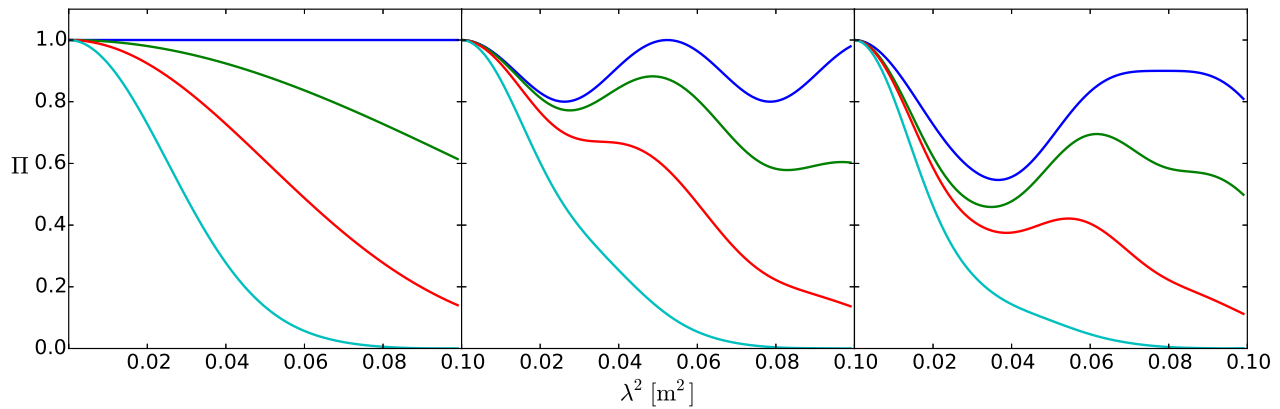}
\caption{Obtained polarization structure $\Pi(\lambda^2)$ for a one component model (left), two
  component model with the secondary component carrying 10\% of the total flux density and 40 rad/m$^2$
  separation from primary component (middle) and three component model with first secondary
  component carrying 20\% of the total flux density and 40 rad/m$^2$ away and the third component carrying
  10\% and 60 rad/m$^2$ separated from the primary component (right). Each model is convolved with a
  Gaussian of width $\sigma_\mathrm{disp}=0$ (blue), $\sigma_\mathrm{disp}=5$~rad/m$^2$ (green),
  $\sigma_\mathrm{disp}=10$~rad/m$^2$\ (red) and $\sigma_\mathrm{disp}=20$~rad/m$^2$ (cyan).}
\label{fig:toydep1}\end{figure*}

This result then led us to re-examine the links between the FD distribution and depolarization,
to isolate that feature of the FD distribution that is most strongly causing depolarization, and
then to construct a further FD parameter that is, finally, clearly associated with the presence of
intervening systems in our sample.  This is the subject of the next two Sections~\ref{sec:toymodel}
and \ref{sec:intervener}.  Furthermore, we are able to argue that the overall structure of the FD
distribution examined in this section is actually likely reflecting other properties of the
sources. This is briefly examined in Section~\ref{sec:intrinsic}.
\newline

\subsection{The connection between FD distribution and depolarization}\label{sec:toymodel}

As commented at the end of the previous section, we found a surprising disconnect between the
evident richness of structure in the FD distribution $F(\phi)$ and the presence of depolarization signatures
in the overall polarization spectrum. The richness of the polarization structure was also surprising.  As
discussed already, we might expect to see simple monotonically decreasing polarization with
increasing wavelength (for a simple Gaussian FD distribution) but instead see a very wide range of
behavior (as in Column~(iii) of Figures~\ref{fig:data1}-\ref{fig:data3}).

The polarized flux density of a given source at a given wavelength represents the vector sum of the
complex representations of the different $\phi$ components, each of which rotates at a speed (in
$\lambda^2$ space) that is proportional to $\phi$. It is worth distinguishing between the effects on
$P(\lambda^2)$ of a few components in $F(\phi)$ that are widely separated in $\phi$, referred to as
``gross structure'' in the following, and the effects of very closely spaced, or continuous,
structure in $F(\phi)$, as for example in the Gaussian width of a particular component, referred to
as ``fine structure''.  The former causes large variations in the polarized flux density with
wavelength as the small number of polars rotate around each other producing an oscillatory behavior
in the total amplitude, i.e. in the polarized flux density.  In contrast, the fine structure in
$\phi$ within a feature produces a more steady decrease in polarization as the continuous
distribution of Faraday Rotation causes a progressive cancellation of polarized flux.

The FD distributions of many sources in Figure~\ref{fig:data1}-\ref{fig:data3} show multiple
discrete components, i.e. gross structure, and the parameters which we constructed in the
previous section, including our own subjective ranking, were based primarily on the presence of
these multiple discrete components rather than the width of individual components.

Further evidence for the distinction between multiple components and the widths of them
(i.e. between gross and fine structure) comes from comparison of the $\sigma_\mathrm{FD}$ values to
the parameters obtained from the depolarization curves (i.e. $\sigma_\mathrm{Burn}$ or
$\sigma'_\mathrm{Burn}$). Although we would have expected those parameters to trace the same
physical quantity, i.e. the dispersion of the FD distribution, we obtain $\sigma_\mathrm{FD}$ to be
an order of magnitude larger than the depolarization parameters. This could be explained by arguing
that $\sigma_\mathrm{FD}$ describes gross structures while $\sigma_\mathrm{Burn}$ and
$\sigma'_\mathrm{Burn}$ describe fine structures. The $\sigma_\mathrm{PC}$ parameter is sensitive to
these fine structure effects because it measures the widths of the primary components. We will show
later that they are of similar size as $\sigma_\mathrm{Burn}$ and $\sigma'_\mathrm{Burn}$.

\begin{figure*}
  \centering\includegraphics[width=0.95\textwidth]{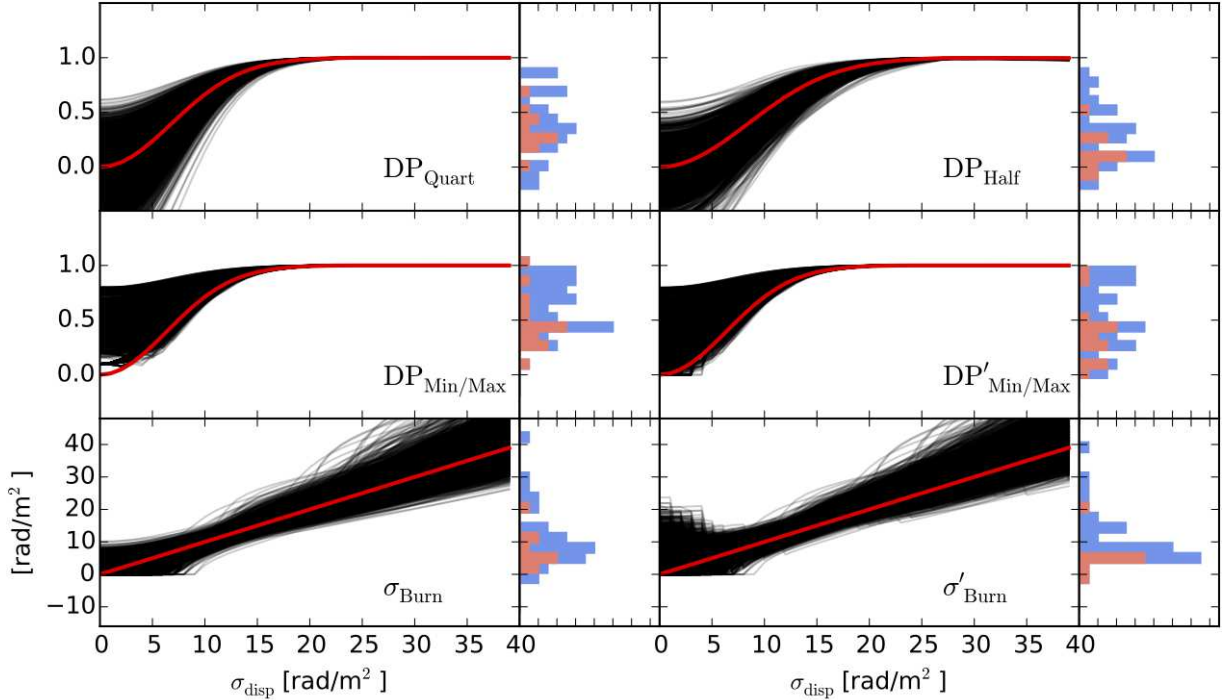}
  \caption{6000 one, two or three component Faraday screen models have been implemented by varying
    relative Faraday Depth, size and initial angle of the components and each model is
    convolved with $\sigma_\mathrm{disp}$ between 0 and 40~rad/m$^2$. The obtained
    depolarization parameter DP$_\mathrm{Quart}$ (top left), DP$_\mathrm{Half}$ (top
    right), DP$_\mathrm{Min/Max}$ (middle left), DP$'_\mathrm{Min/Max}$ (middle right),
    $\sigma_\mathrm{Burn}$ (bottom left) and $\sigma'_\mathrm{Burn}$ (bottom right) for each model
    is represented by the black lines. The red line represents the one component model. The
    histograms on the right show the respective depolarization parameter distribution of objects
    without (red) and with (blue) interveners in the sample. The histograms are stacked.}
  \label{fig:toydep3}\end{figure*}

We explore the effects of both discrete components and the Gaussian width on the polarization
$P(\lambda)$ using a simple toy model. The toy model consists of $N=1000$ FD cells which are
each associated with $\phi_i$ according to an underlying FD distribution, assuming that all cells have
the same initial phase $\psi$ and also assuming unit total flux density, $I$, so that
$P(\lambda) = \Pi(\lambda)$. The polarization $P(\lambda^2)$ is then determined by
\begin{equation}
  P(\lambda^2)=\sqrt{U(\lambda^2)^2+Q(\lambda^2)^2}
\end{equation}
where
\begin{eqnarray}
  U(\lambda^2)&=&\frac{1}{N}\sum^N_i \sin(2\phi_i\lambda^2) \\
  Q(\lambda^2)&=&\frac{1}{N}\sum^N_i \cos(2\phi_i\lambda^2) \;.
\end{eqnarray}

Figure \ref{fig:toydep1} shows the resulting depolarization structure of a few simple models. The
left panel corresponds to a model with one $\phi$ component, the middle panel to a model with two
components where the secondary component carries 10\% of the total flux density. and the right panel to a
model with three components where the first secondary component carries 20\%, and a second 10\%, of
the total flux density. All $\phi$ components have been convolved with a Gaussian of different widths in
$\phi$, namely $\sigma_\mathrm{disp}=0$ (blue), $\sigma_\mathrm{disp}=5$~rad/m$^2$ (green),
$\sigma_\mathrm{disp}=10$~rad/m$^2$ (red) and $\sigma_\mathrm{disp}=20$~rad/m$^2$ (cyan),
respectively.

We see that highly non-monotonic and complex polarization structure is produced by adding just a few
additional components. However, the overall depolarization is mainly driven by the convolution with
the Gaussian rather than by these multiple components. Multiple discrete components in contrast tend
to add oscillatory features to the polarization structure.  The small amplitude of the oscillations
reflects the small fractions of the flux density in the secondary components. In contrast, the convolution
leads to phase dispersions affecting all of the polarized flux density.

We can now return to the different parameterizations of the depolarization introduced in
Section~\ref{sec:dfdep} and examine which of them best reflects the fine structure
$\sigma_\mathrm{disp}$, recalling that these were correlated to different degrees with the presence
of MgII intervening absorption. We use the same type of simple models, but now construct 6000 models
by varying the positions of the secondary component(s) relative to the primary component (between
zero and 200 rad/$m^2$), by varying the combined relative flux density contribution of the secondary
component(s) (up to 40\% of the total) and also by varying the initial phases $\psi$ of the
secondary component(s) relative to the primary component, and to each other (between 0 and
$\pi$). These variations reflect what we see in the FD distributions of our sample quasars
(cf. $\sigma_\mathrm{FD}$ and $C$ in Figure~\ref{fig:fdist}). Furthermore each of these 6000 models
is convolved with a range of $\sigma_\mathrm{disp}$ between 0 and 40~rad/m$^2$.  The obtained DP
parameters of those models are shown in the different panels of Figure~\ref{fig:toydep3}. As with
the real data, it was not always possible to fit $\sigma_\mathrm{Burn}$ (bottom left) and
$\sigma'_\mathrm{Burn}$ (bottom right) with physically meaningful values.  The red lines in each
panel shows the behavior of the single component model. As a comparison, the distribution of the
corresponding DP parameters in our observed quasar sample are represented by the histograms on the
right-hand axis.

As discussed above, the feature in the FD distribution which most drives depolarization is
$\sigma_\mathrm{disp}$. Hence for our purposes a ``good'' DP parameter should be one which well
traces $\sigma_\mathrm{disp}$.  Figure \ref{fig:toydep3} shows the relation of the previously
defined DP parameters to $\sigma_\mathrm{disp}$ for the simple one component model (red
line) and as a comparison for more complex models with two or three components (black lines). It
shows how the appearance of additional components can bias DP and blur the relation between
$\sigma_\mathrm{disp}$ and DP.  We see that DP$_\mathrm{Quart}$ (top left) and DP$_\mathrm{Half}$
(top right) can be both increased and decreased by the presence of secondary components while
DP$_\mathrm{Min/Max}$ (middle left) and DP$'_\mathrm{Min/Max}$ (middle right) mostly tend to be
increased.

Both $\sigma_\mathrm{Burn}$ (bottom left) and $\sigma'_\mathrm{Burn}$ (bottom right) work
surprisingly well in recovering an estimate of $\sigma_\mathrm{disp}$, despite the difficulties of
fitting the Burn model due to the non monotonic structure of $\Pi(\lambda^2)$.  For
DP$_\mathrm{Quart}$ and DP$_\mathrm{Half}$ most of the objects in our sample are in that region
where secondary components can strongly affect them.  As opposed to DP$_\mathrm{Quart}$ or
DP$_\mathrm{Half}$, we see that for DP$_\mathrm{Min/Max}$ and DP$'_\mathrm{Min/Max}$ a fair portion
of the sample have values close to 1, where the value is more robust against additional components.

This analysis, based on a simple toy model, therefore offers a possible explanation as to why we got
different KS significances for the different DP parameters in Section~\ref{sec:depol}.  Recall that
we saw the strongest correlations with the presence of intervening absorption with
DP$_\mathrm{Min/Max}$ and DP$'_\mathrm{Min/Max}$, and also, when measurable, with
$\sigma_\mathrm{Burn}$ and especially $\sigma'_\mathrm{Burn}$.  We would have seen this behavior if
the presence of MgII absorption is associated with the fine structure $\sigma_\mathrm{disp}$ rather
than the presence of the multiple components which dominate the visual impression of the FD
distributions in Figures~\ref{fig:data1}-\ref{fig:data3}.

We therefore suggest that intervening material is primarily responsible for broadening the FD
distribution $F(\phi)$ rather than the presence of the multiple discrete components. We will test this in the
next Section~\ref{sec:intervener} and provide a further argument in favor of this idea in the subsequent
Section~\ref{sec:intrinsic}.

\subsection{Intervener Effects in Faraday Depth}\label{sec:intervener}

In the previous section we postulated that intervening material, traced by the MgII absorption,
affects the observed FD distribution by means of a convolution effect. $\sigma_\mathrm{PC}$,
introduced in Section~\ref{sec:deffddist}, is a parameter constructed to be insensitive to the
appearance of multiple components.

\begin{figure}
  \centering\includegraphics[width=0.45\textwidth]{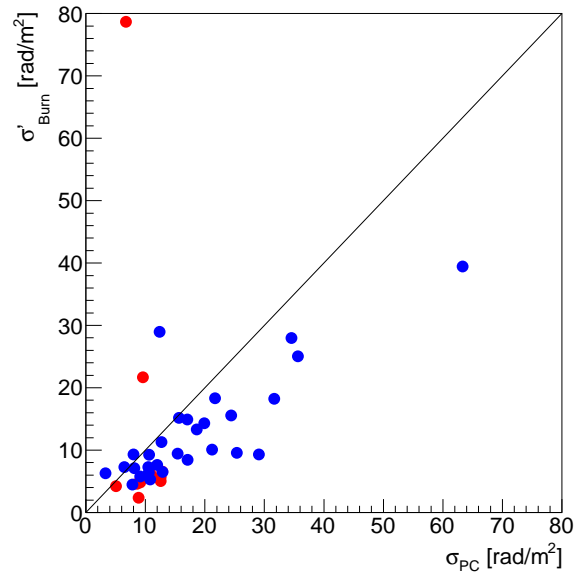}
  \caption{Comparison between the depolarization parameter $\sigma'_\mathrm{Burn}$ and the standard
    deviation of the primary component $\sigma_\mathrm{PC}$ for objects without (red) and with
    (blue) interveners. 10 objects for which the $\sigma'_\mathrm{Burn}$ fit failed are missing.}
  \label{fig:sigmadispcor}\end{figure}

In Figure \ref{fig:sigmadispcor} we compare the $\sigma_\mathrm{PC}$ to $\sigma'_\mathrm{Burn}$. The
$\sigma_\mathrm{RMSF}$ in the VLA data is around 17 rad/m$^2$ and is around 24 rad/m$^2$ for ATCA.
Most objects have $\sigma_\mathrm{PC}$ smaller than that, i.e. those sources are barely resolved in
$\phi$ space, and the values should be interpreted with some caution.  Nevertheless, and despite the
difficulties in obtaining fits to $\Pi(\lambda^2)$ with the Burn model, we see a reasonable overall
correlation between $\sigma_\mathrm{PC}$ and $\sigma'_\mathrm{Burn}$.

\begin{figure}
  \centering\includegraphics[width=0.45\textwidth]{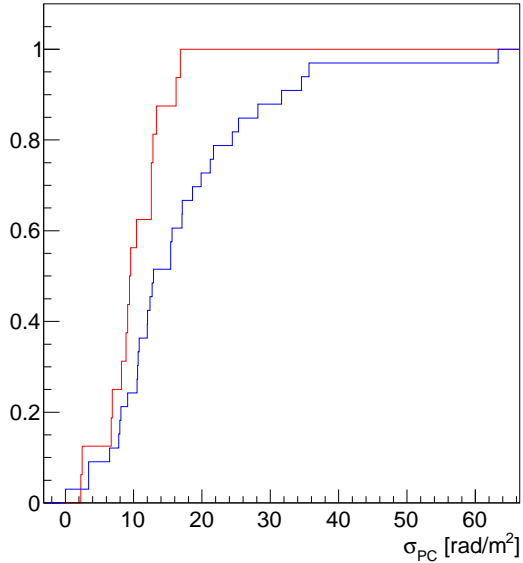}
  \caption{EDF of the depolarization parameter $\sigma_\mathrm{PC}$ for objects without (red) and
    with (blue) interveners. The one-tailed KS test yields $p=3.5\%$.}
  \label{fig:sigmadispcum}\end{figure}

If we now construct our usual KS test between absorption and $\sigma_\mathrm{PC}$, i.e. a one-tailed
test with the null hypothesis that objects with MgII absorption along the lines of sight are drawn
from the same distribution of $\sigma_\mathrm{PC}$ as objects with clean lines of sight or that
objects with clean lines of sight have enhanced $\sigma_\mathrm{PC}$, we can reject the null
hypothesis with $p=3.5\%$ (Figure~\ref{fig:sigmadispcum}). When re-scrambling objects with and
without MgII absorption along the lines of sight as in Section~\ref{sec:rm} $p=3.5\%$ or smaller
happens in 1.9\% of the realizations.

To assess the significance of this result, it should be borne in mind
that most of the objects were not resolved in $\phi$ space. It is noticeable that
\textit{all} of the quasars with $\sigma_\mathrm{PC}$ significantly larger than
$\sigma_\mathrm{RMSF}$ have intervening MgII absorption.

When we use the fractional $q$ and $u$ for RM Synthesis we obtain here an even more significant
p-value, with $p=0.09\%$. Sources with spectral indices different than zero, i.e. sources which vary
along the observed frequency, will have associated variations also in the polarized flux density,
i.e. in $Q$ and $U$, and this will affect the FD distribution. Since these variations are normally
gradual with frequency, these will rather affect parameters like $\sigma_\mathrm{PC}$ than adding
new secondary components. This then would blur the KS Test with
$\sigma_\mathrm{PC}$. Using fractional $q$ and $u$ is an attempt to take out this effect and indeed
it yields a stronger result.

The strong association of both, depolarization and $\sigma_\mathrm{PC}$, with intervening material
indicates that intrinsic effects within the sources do not contribute much to the fine structure of
the Faraday Depth within each components, i.e. $\sigma_\mathrm{disp}$. This could be due to
different properties of the magnetized plasma within sources compared to that in the intervening
systems traced by MgII absorption.  However, redshift effects should also be considered here.  A
given dispersion in the rest-frame $F(\phi)$ at some redshift $z$ will produce an observed
$\sigma_\mathrm{disp}$ that scales as $(1+z)^{-2}$.  Since the quasars are generally at
significantly higher redshifts than the intervening absorbing systems, this could explain why the
observed $\sigma_\mathrm{disp}$ is dominated by the intervening systems.

\subsection{Intrinsic effects in Faraday Depth}\label{sec:intrinsic}

We have argued above that the presence of multiple discrete components in the FD distributions
$F(\phi)$ of our sources (to which the parameters we introduced in Section~\ref{sec:nullresult} were
particularly sensitive) is not associated with the presence of intervening MgII absorption. In
contrast, intervening material is clearly associated with the broadening of the FD
distribution. This raises the possibility that the multiple components in $F(\phi)$ arise actually
intrinsically to the radio sources.

That this is likely the case is indicated when we consider the initial phases $\psi$ of the
components. This is shown on Figure \ref{fig:data1}-\ref{fig:data3} (Column~(vi)).  We always
see that if a secondary component is present, then the initial phase of that secondary component is
different than the initial phase of the primary component.  In other words, the cusps in the
radial $\mathbf{F}$ plots (in Column~(vi)), that correspond to the peaks in the $F(\phi)$ plots
in Column~(v), lie at different azimuthal angles representing the initial phase $\psi$.

An important point is that source components with different initial phase $\psi$ \textit{must} come
from spatially distinct emission regions, either in the plane of the sky, or along the line of
sight.  This is because $\psi$ represents the phase of emission before any Faraday Rotation.  It is
then quite reasonable to imagine that these different emission regions have associated with them
different intrinsic $\phi$. This would naturally produce the multiple $\phi$ components, each with
distinct $\phi$ and $\psi$, in the overall FD distribution, as observed.

This argument above is however not watertight: It could also be imagined that different source
regions (with different $\psi$) lying behind different parts of an intervening system would also
suffer vastly different amounts of Faraday Rotation.  We could for instance imagine a chequer-board
intervening screen with just two values of intervening $\phi$, $\phi_1$ and $\phi_2$, which lies in
front of a source in which $\psi$ varies spatially.  We would then observe a Faraday Depth with two
separated peaks $\phi_1$ and $\phi_2$, each with a $\psi$ that reflected the distribution of $\psi$
of the source regions behind the $\phi_1$ and $\phi_2$ cells of the intervening screen.  This would
also produce two distinct components in $F(\phi)$ with different $\psi$, with all of the Faraday
Rotation coming from the intervening system. However, in this case, we would expect to see a
correlation between the presence of MgII and the parameters introduced earlier that are sensitive to
the presence of these multiple components in the FD distribution.  This is not seen, suggesting that
indeed, the different $\phi$ of the multiple components originate intrinsically to the source.

Also the fact, that we do not see a single example in which two or more cusps have the same $\psi$
which would happen e.g. if a chequer-board intervening screen as described before lies in front of a
source with constant $\psi$, should be taken as an indication that distinct components in the FD
distribution arise from intrinsic effects.

We conclude therefore that the gross structure complexity and multiple distinct components in
$F(\phi)$ that are seen in many sources are produced by effects intrinsic to the sources, while
intervening material introduces a small but pervasive broadening of the FD distribution.  We explore
further the implications of the latter effect in the final Discussion section immediately following.
Further exploration of the properties of the background sources from QU-fitting (\cite{sul12},
\cite{sun15}) and analysis of the shape of the $\mathbf{F}$ plots is beyond the scope of the current
work but we plan to develop this further in the future. The results of the QU-fitting will also be
very interesting to be compared with $\sigma_\mathrm{PC}$.

\section{Discussion}\label{sec:discussion}

In this section we will try to be more quantitative about the size of the Faraday Effects that are
associated with the presence of intervening material, as revealed by MgII absorption. This will
enable us to better characterize the physical properties of this material.

We estimate the dispersion of $\phi$ introduced by intervening systems. The largest separation
in the $\sigma_\mathrm{PC}$ empirical distribution functions for sources with and without
interveners is observed at $\sigma_\mathrm{PC}=16$~rad/m$^2$.  As noted earlier, this value should
be treated with caution because the $\phi$ resolution limit of our experiment is roughly
$\sigma_\mathrm{RMSF}=17$~rad/m$^2$ for VLA and $\sigma_\mathrm{RMSF}=24$~rad/m$^2$ for ATCA. The
difficulties of the RMSF can to a certain degree be circumvented by simply subtracting in quadrature
the median $\sigma_\mathrm{PC}$ of sources with and without absorbers (since the RMSF has been
subtracted from both $\sigma_\mathrm{fit}$ equally). This yields an estimate for the excess
$\sigma_\mathrm{PC} \approx 8.6$~rad/m$^2$.

The maximum separation seen in the $\sigma'_\mathrm{Burn}$ distribution is at
$\sigma'_\mathrm{Burn}=7$ rad/m$^2$. But also here this value should be treated with caution because
of the non-monotonic structure in $\Pi(\lambda^2)$ and the rather \textit{ad hoc} approach to
choosing the wavelength range for the fit (only long-wards of the peak in polarization). Likewise,
subtracting the median $\sigma'_\mathrm{Burn}$ values of the distribution with and without absorbers
in quadrature gives an estimate for the excess $\sigma'_\mathrm{Burn} \approx 8$~rad/m$^2$.

$\sigma_\mathrm{PC}$ and $\sigma'_\mathrm{Burn}$ attempt to trace $\sigma_\mathrm{disp}$ by
different approaches. The fact that their values are similar gives some confidence that the proposed
scenario is self-consistent.

From the current work we will use a value of order 10 rad/m$^2$ as a crude order of magnitude estimate of the FD
dispersion that is associated with intervening systems. This is an estimate of the range of the
observed $\phi$ \textit{within} (i.e. across the face of) a given intervening system.  To convert to
the rest-frame of the absorber, we must multiply by $(1+z)^2$, i.e. by a factor of order 2 for our
median absorber redshift of $z \approx 0.5$, yielding of order
$\sigma_\mathrm{int,0} \approx 20$~rad/m$^2$. This value is of the same order of magnitude as the
spread in Galactic RMs $\sigma_\mathrm{MW} \approx 8$~rad/m$^2$ as estimated by \cite{schn10}. Other
works have claimed $\sigma_\mathrm{LMC} \approx 81$~rad/m$^2$ for the Large Magellanic Cloud
(\cite{gae05}) and $\sigma_\mathrm{SMC} \approx 40$~rad/m$^2$ for the Small Magellanic Cloud
(\cite{mao08}).

As in \cite{bern08} we can crudely associate a $\phi$ with an estimate of magnetic field strength
using an estimate of the free electron column density in the interveners based on the known MgII
column density. Assuming a compact absorber system at redshift $z$ and the total magnetic field
strength to be $B=\sqrt{3}B_\parallel$ this reformulates equation~\ref{eq:RM} as (see \cite{bern08})
\begin{equation}
B=6.6\times 10^{12} \frac{(1+z)^2}{N_\mathrm{e}} \phi
\end{equation}
where $B$ is in units of G, $N_\mathrm{e}$ is the ion column density in units of cm$^{-2}$ and
$\phi$ is the Faraday Depth in units of rad/m$^2$. This estimate ignores any field reversals within
the system. Moreover, assuming neutral-hydrogen column density of $N(HI)\approx10^{19}$~cm$^2$
(\cite{rao06}) and a hydrogen ionization fraction of $\bar{x}\approx0.90$ (\cite{pro06},
\cite{per07}) we estimate $N_\mathrm{e}\approx9\times10^{19}$~cm$^2$.  Values of
$\sigma_\mathrm{int}$ of order 10~rad/m$^2$ are associated with random fields of order
$\sigma_B\approx\sqrt{m}\times3\,\mu$G, where $m$ is the typical number of reversals within the
intervening system along a given one-dimensional line of sight. Since the physical properties of the
intervening absorption systems are barely constrained, even order of magnitude estimates of $m$
would be unreliable at this stage.

Recently \cite{rie16} have investigated small-scale dynamo effects in high redshift galaxies
performing magneto-hydrodynamic simulations which enables them to predict the strength and
turbulence of magnetic fields in the circumgalactic medium. A detailed comparison of their
predictions with our observations is beyond the scope of this paper, but will be presented
elsewhere.

Finally we remark that our results emphasize that the clearest signatures of the presence of
intervening material are to be seen in the dispersion of Faraday Depth $\sigma_\mathrm{disp}$ within
the spatial extent of a given individual source, as traced by our $\sigma_\mathrm{PC}$ and
depolarization parameters, rather than in the overall Faraday Depth, that is traced by
$\phi_\mathrm{max}$ or RM (\cite{bern08}), which instead are measuring uniform effects across the
whole source. This should not be surprising. Because of orientation effects, the enhancement of
$\phi_\mathrm{max}$ for sources with MgII absorption lines compared to clean sources can only be
made statistically, i.e. by comparing the widths of the $\phi_\mathrm{max}$ or RM distributions,
while $\sigma_\mathrm{PC}$ as well as the depolarization parameters are a direct measure on a source
by source basis. Put another way, the signal of interveners will be seen in the first moment of the
$\sigma_\mathrm{disp}$ distribution, but only in the second moment of the $\phi_\mathrm{max}$ or RM
distribution.

We also remark on the fact that most of the $F(\phi)$ distributions are barely resolved by the RMSF
of the instrumental set up, especially for the sources without intervening systems. This suggests
that further improvements in this through the extension of the wavelength baseline in RM Synthesis
may be useful.

\section{Summary}\label{sec:summary}

We have presented new continuous radio polarization measurements taken with the VLA and ATCA of 49
unresolved quasars for which we also have a census of intervening MgII absorption systems, down to
0.1~\AA~rest frame equivalent width.  We apply RM Synthesis on the radio data and compile a set of
high quality $F(\phi)$ FD distributions.  The $F(\phi)$ exhibit a rich diversity in
structure and complexity.

In essentially all objects there is a pronounced, roughly Gaussian, primary component in $F(\phi)$.
About two thirds of the sample also exhibit additional secondary components, but these display a range of
structure. The overall polarization of the sources $P(\lambda)$ are also strikingly varied.

We have investigated the connections between features in the $F(\phi)$ distributions and the
polarization $P(\lambda)$ and correlations between these and the presence of interveners, and reach
the following main conclusions:
 
\begin{itemize}
\item We compare our completely independent radio data with the RM data used in \cite{bern08} and
  can, in the main, confirm the reliability of the latter. However, with our substantially smaller
  sample we only can recover a marginal result for the connection between peak $\phi_\mathrm{max}$
  and the presence of MgII absorption in the optical spectrum.

\item We demonstrate, however, a strong connection between the presence of intervening absorption and the
  depolarization of the source, obtaining $p\approx1\%$ with a one-tailed KS test for three
  different parameters describing depolarization. This result firmly establishes the connection
  between radio properties and intervening absorption and affirms the inhomogeneous nature of the
  intervener screens.
  
\item Surprisingly, given the strong association with depolarization, we do not see significant
  correlations between the gross structure complexity of $F(\phi)$ and the presence of
  interveners, or indeed with our measures of depolarization.  Instead, analysis of the initial
  phases of the different $\phi$ components, suggests that the most visible structure in $F(\phi)$ that is
  caused by multiple discrete components, actually stems from effects intrinsic to the background
  radio sources.
  
\item In contrast, the effect of the intervening systems is manifested by a systematic broadening of
  the overall $F(\phi)$.  Parameters describing this broadening are correlated with the presence of
  intervening absorption ($p\approx5\%$). Furthermore, we also show that it is this fine structure
  broadening of components in $F(\phi)$, rather than the presence of multiple components, that is
  primarily responsible for the observed depolarization, tying this result back to the strong
  connection with depolarization.
  
\item We estimate that the typical effect of intervening MgII absorber systems (which generally have
  $0.4 < z < 1.4$) is to produce a broadening of the individual components in $F(\phi)$ of order
  10~rad/m$^2$.  Similar estimates come from fitting models to the depolarization spectrum. This can
  be associated with random fields of order $\sigma_B\approx\sqrt{m}\times3\,\mu$G, where $m$ is the
  number of field reversals along the line of sight.
  
\end{itemize}

Our new findings considerably strengthen the evidence that intervening MgII absorption systems have
detectable signatures on the polarization properties of background quasars, and that these can be
used to probe the magnetic field structure of these systems, which are known to be the outflow
regions of star forming galaxies at intermediate redshifts.

\acknowledgments{We are grateful to the anonymous referee for their comments which drew our
  attention to the possible Galactic effects. We thank Philip Schmidt (MPIfR) for valuable help
  during the calibration, imaging and RM Synthesis of the VLA data. We also thank Aritra Basu for
  refereeing this report internally at MPIfR. This research has been supported by the Swiss National
  Science Foundation. This work was supported by the Australian Research Council through grants
  FL100100114 and FS100100033, and by the National Sciences and Engineering Research Council of
  Canada through grant RGPIN-2015-05948. The Dunlap Institute is funded through an endowment
  established by the David Dunlap family and the University of Toronto. SPO acknowledges support
  from UNAM through the PAPIIT project IA103416.  National Radio Astronomy Observatory is a facility
  of the National Science Foundation operated under cooperative agreement by Associated
  Universities, Inc. The Australia Telescope Compact Array is part of the Australia Telescope
  National Facility which is funded by the Australian Government for operation as a National
  Facility managed by CSIRO.}


\bibliography{kim}

\end{document}